\documentclass[10pt,journal, oneside, twocolumn]{IEEEtran}

\usepackage{amsfonts}
\usepackage{diagbox}
\usepackage[normalem]{ulem}
  \usepackage{graphicx}

  \usepackage{algorithm}
  \usepackage{algorithmic}
  \usepackage{multirow}
\usepackage{color}
  \usepackage{threeparttable}
  \usepackage{graphicx}
\usepackage{geometry}
  \usepackage{stfloats}
  \usepackage{booktabs}
  \usepackage{threeparttable}
  \usepackage{textcomp}
  \usepackage{epstopdf}
  \usepackage[noadjust]{cite}
  \usepackage{array}
  \usepackage{ulem}
  \usepackage{rotfloat}
  \usepackage{longtable,booktabs}
  \usepackage{supertabular,booktabs}
  \usepackage{amsmath}
  \usepackage{mathrsfs}
  \usepackage{changebar}
  \usepackage{lipsum}
  \usepackage{setspace}
  \usepackage{changebar}
  \usepackage{float}
  \usepackage[caption=false]{subfig}

\newtheorem{remark}{Remark}

\newboolean{HAVEAAA}
\setboolean{HAVEAAA}{false}

\begin{document}
\begin{spacing}{1.0}

\title{H-AP Deployment for Joint Wireless Information and Energy Transfer in Smart Cities}
\author{Yizhe Zhao,~\IEEEmembership{Student Member,~IEEE}, Duohua Wang, Jie Hu,~\IEEEmembership{Member,~IEEE}, Kun~Yang,~\IEEEmembership{Senior Member,~IEEE}
\thanks{Copyright (c) 2015 IEEE. Personal use of this material is permitted. However, permission to use this material for any other purposes must be obtained from the IEEE by sending a request to pubs-permissions@ieee.org.}
\thanks{Yizhe Zhao and Jie Hu are with the School of Information and Communication Engineering, University of Electronic Science and Technology of China, Chengdu, 611731, China, email: yzzhao@std.uestc.edu.cn, hujie@uestc.edu.cn. Jie Hu is the corresponding author.}
\thanks{Duohua Wang is with ZTE corporation, Xi'an 710065, China, email: wang.duohua3@zte.com.cn.}
\thanks{Kun Yang is with the School of Computer Science and Electronic Engineering, University of Essex, Colchester, CO4 3SQ, U.K., and also with the School of Information and Communication Engineering, University of Electronic Science and Technology of China, Chengdu 611731, China, e-mail: kunyang@essex.ac.uk.}
\thanks{The financial support of National Natural Science Foundation of China (NSFC), Grant No. 61601097, Grant No. U1705263 and Grant No. 61620106011 as well as that of University of Electronic Science and Technology of China, No. ZYGX2016Z011, are gratefully acknowledged. We would also like to thankfully acknowledge the financial support of the 111 project (NO.B14039) and that of the ZTE Corporation.}}

\maketitle

\begin{abstract}
With the wireless communication being more various in the future, it's becoming challenging to prolong the lifetime of many battery powered devices, since frequently replacing their batteries is a cumbersome job. An hybrid access point (H-AP) is capable of simultaneously operating wireless information transfer (WIT) and wireless energy transfer (WET) by exploiting the radio frequency (RF) signals. By jointly considering both the mobility of the user and the popularity of the sites, we focus on the design of the  H-AP's deployment scheme. Specifically, a mobility model of the grid based city streets is exploited for characterising the users' movements. Based on this mobility model, the impact of the deployment site's popularity on the WIT and WET efficiencies is firstly analysed.  Then, an H-AP deployment scheme for striking a balance between the WIT and the WET efficiencies is proposed, which is regarded as the B-deployment scheme.  The simulation results demonstrate that the B-deployment scheme is more flexible for satisfying diverse requirement of the WIT and WET efficiencies.
\end{abstract}

\begin{IEEEkeywords}
wireless information transfer, wireless energy transfer, mobile social networks, H-AP deployment
\end{IEEEkeywords}

\IEEEpeerreviewmaketitle

\section{INTRODUCTION}

With the development of wireless communication technologies, wearable  and pocket communication devices such as E-glasses, E-watches, smart belts and \textit{etc.}, have penetrated into our daily lives. All these communication devices require to exchange vital information with central servers, such as the users' physical condition, the location information and \textit{etc}.

\begin{figure}
  \centering

  \includegraphics[width=3.5in]{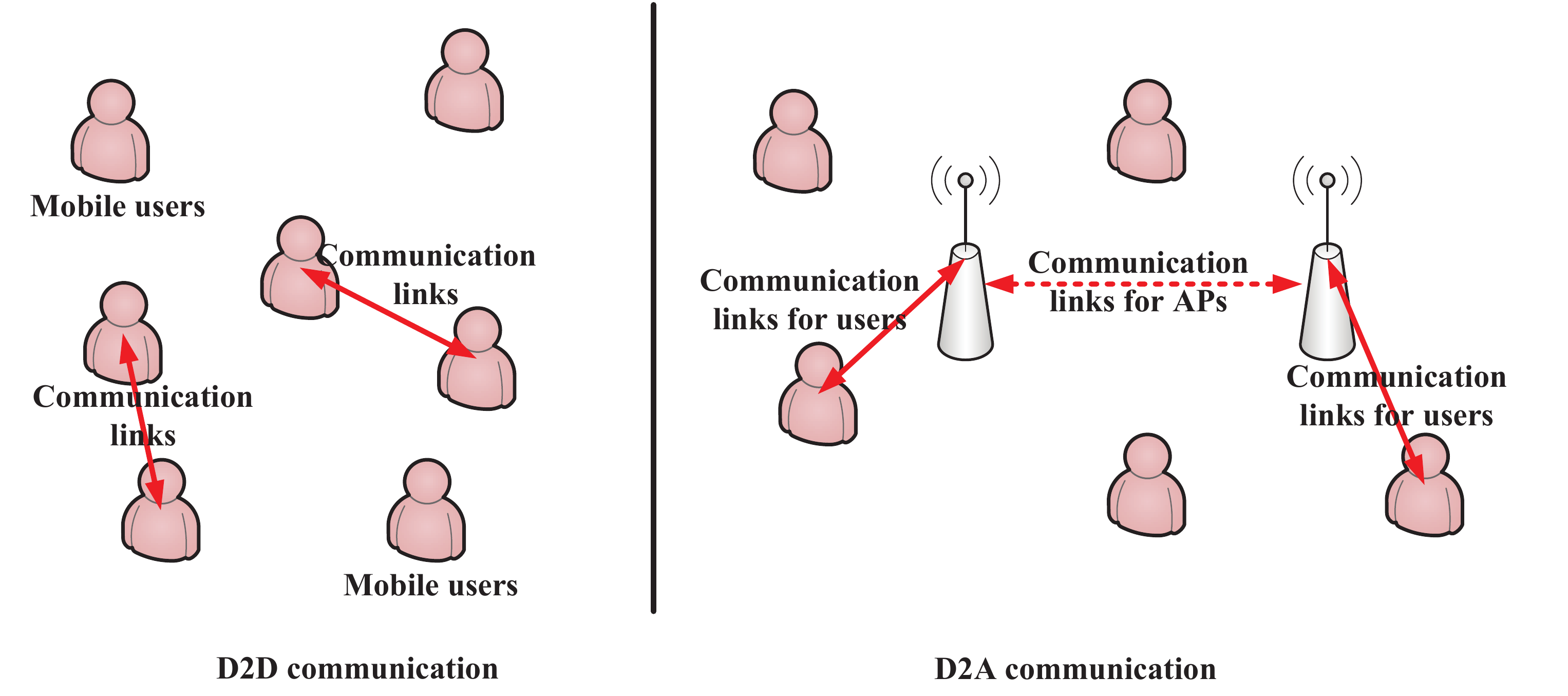}
  \caption{Communication types in MSN}\label{MSN_type}
\end{figure}

Delay tolerant networks (DTNs) \cite{1717} have been proposed for realising communication in intermittent networks. In DTNs, a direct communication link between any pair of nodes is not always available, due to extreme terrestrial environments or movements of the communication nodes. In order to improve the performance of the information delivery, Akhtar \textit{et al.} \cite{1111} has proposed to exploit the users' social properties for improving the information delivery performance of the DTNs, which yields a new concept of mobile social network (MSN). Nikolaos \textit{et.al.} \cite{2828} proposed the architecture and social properties of MSN, while also threw some key research challenges. There are two communication types in MSNs, namely device to device (D2D) communication \cite{3131} and device to access point (D2A) communication, as potrayed in Fig.\ref{MSN_type}. In the D2D communication, a pair of different users may establish a direct communication link between their devices without the aid of any centralised infrastructure, if they enter each other's transmission range. For example, Jie \textit{et al.} \cite{1818} has proposed a content dissemination scheme by exploiting the D2D communications among the users. By contrast, in the D2A communication, a user may communicate with an centralised AP, if it enters the AP's transmission range. For instance,  Zhao \textit{et al.}  \cite{1212} has studied a joint AP deployment and routing scheme for efficiently delivering information to the users in MSNs. Later on, Ying \textit{et al.} \cite{1313} have proposed another AP deployment  scheme by exploiting the social popularity of specific locations. Furthermore, Fan  \textit{et al.} \cite{1414} \cite{1515} have developed a joint scheme for both the AP's deployment and the storage allocation in order to efficiently deliver popular content towards the users. In contrast to the conventional high-power cellular communication, the APs in the MSN often operates at a low transmit power. As a result, the transmission range  of the APs is limited. Hence, users have to enter the transmission range of the APs so as to download important information to their wearable devices. Therefore, users' mobility largely determines the communication performance of the MSN.

Furthermore, due to their limited size, wearable devices are normally equipped with batteries having limited capacity. The batteries may quickly drain for the sake of powering complex signal processing and sensing tasks. Yang \textit{et.al.} \cite{2929} proposed an approach to offload services for resource-constrained mobile devices, while the resource management can also be operated on the cloud to reduce the energy consumption \cite{3232}. However, this is not an ideal solution as some tasks are simply not suitable to run on servers and task offloading also incurs extra energy consumption. Therefore, it is essential to incorporate the function of wireless energy transfer (WET) to the conventional communication oriented APs, which yields the hybrid access points (H-APs). Many radio frequency (RF) signals based WET techniques have been studied in different OSI layers \cite{3030}, such as the energy beamforming \cite{555} and channel coding \cite{222} in the physical layer, the network architecture \cite{333} and the MAC protocol design \cite{444}\cite{777} as well as the resource allocation \cite{666}  in the network and MAC layer. Moreover, coordinating both the WIT and WET in the same RF spectral band has attracted much attention from both the academia and industry, which triggers substantial works in simultaneously wireless information and power transfer (SWIPT) and wireless powered communication (WPC). All these related works \cite{888,999,1010} aim for striking a balance between the WET and the WIT performance by exploiting the hot communication techniques, such as the massive MIMO technique \cite{1010}, full duplex technique \cite{2424} and channel estimation technique \cite{888}.

However, many of the energy concerns focus in the future cellular networks \cite{3333}, while few existing works study the WET in the scenario of MSNs, where both the WIT and WET performance is dominated by the users' mobility patterns. Fan \textit{et al.} \cite{1616} have firstly proposed to exploit the so-called enhanced throwboxes for realising both the WIT and WET in the MSN. However, their model is extremely simplified without considering a realistic mobility model, the wireless channel attenuation and the distinctive features of the WIT and the WET. In order to fill this gap, our novel contribution can be summarized as follows
\begin{itemize}
\item A user mobility model  in the city streets is exploited for characterising the users' mobility patterns, in which users move randomly along the city streets and they may randomly opt to turn or to go straight, when they reach a crossroad.
\item A two-dimensional Markov chain is exploited for analysing the seminal metrics of the users' mobility  patterns, including the users' total sojourn duration within the WIT range of the H-AP and that within the WET range.
\item Both the WIT and WET efficiencies of the MSN studied are derived in close-form. Then, three different H-AP deployment schemes are proposed, namely the I-deployment scheme for maximising the WIT efficiency, the E-deployment for maximising the WET efficiency and the B-deployment for striking a balance between the WIT and WET efficiencies. The algorithms for obtaining both the optimal and sub-optimal solutions are developed.
\end{itemize}

The rest of the paper is organized as follows. In section II, both the network model and the users' mobility model are introduced. Then, the users' mobility patterns are analysed by exploiting a two-dimensional Markov chain in Section III. In section IV, three H-AP deployment  schemes having different goals are proposed, whose solution is obtained in Section V. After the simulation results presented in Section VI, our paper is concluded in Section VII.

\section{System Model}
\subsection{Network Model}
We focus our attention on the typical D2A communication in a mobile social network (MSN),  where users are equipped with wearable devices and moving around street blocks. Users may download important information and charge their batteries via RF signals emitted by the H-APs. Due to the limited transmission range of the H-APs, users' mobility and the popularity of the hot-spots (e.g. shopping mall, parks and sport centres) may jointly determine the performance of both the WIT and WET.  As a result, an optimal deployment scheme of the H-AP should consider both the users' mobility and the popularity of the hot-spots.   We assume  $M$ users  in the MSN studied, which are denoted by a set of $\{u_1,\dots,u_m,\dots,u_M\}$. We also assume that $N$ sites in the MSN studied as the candidates for the deployment of the H-APs, which are  denoted by a set of $\{s_1,\dots,s_n,\dots,s_N\}$. Furthermore, we have $K$ H-APs in total pending to be deployed. A vector $\mathbf{y}=\{y_1,\dots,y_n,\dots,y_N\}$  is exploited for characterising a specific deployment scheme of the H-APs, where $y_n=1$ indicates that a H-AP is deployed at the site $s_n$, while $y_n=0$ indicates that no H-AP is deployed at the site $s_n$. Obviously, if we have more H-APs than the number of sites, say $K \geq N$, each site can have a single H-AP to be deployed. By considering the deploying cost, we normally have fewer H-APs than the number of sites, say $K < N$. In order to maximise the benefit from these $K$ H-APs, an optimal deployment scheme is required.

The H-APs are equipped with a pair of radio fronts operating in different spectral bands for facilitating the WIT and the WET, respectively. Specifically, the WET operates in a low frequency band, which may substantially reduce the path loss induced channel attenuation \cite{2525}. By contrast, the WIT operates in a high frequency band in order to obtain a high bandwidth for increasing the throughput.  The energy harvesting circuit and the information reception circuit at a wearable device have diverse requirement on the minimum power of the received RF signal.  For example, the minimum power of the received RF signal at a wearable device can be -80 dBm for successful information recovery. By contrast, the minimum power requirement on the received RF signal is at least -40 dBm for activating the energy harvesting circuit \cite{2727}. As a result, given the same transmit power, the H-AP has distinct WIT range and the WET range. If a H-AP is deployed at the site $s_n$, its WIT range is denoted as $r_n^D$, while its WET range is denoted as $r_n^E$. Since the energy harvesting requires higher received signal power than the information decoding, we normally have $r_n^E < r_n^D$.  Furthermore, homogeneous H-APs are conceived in the MSN studied. As a result, all the H-APs pending to be deployed have an identical WIT range , denoted as $r^D_n=r^D$, and they also have an identical WET range,\footnote{The wireless channel power gain is mainly affected by the path-loss as well as the multipath fading. We assume that the multipath fading has the same statistical properties at each crossroad. Hence, given the information outage probability of the received WIT SNR lower than an identical threshold, the WIT ranges of the H-APs are identical. So are the WET ranges of the H-APs. Therefore, we assume that the WIT and the WET ranges of all the H-APs are identical.} denoted as $r^E_n=r^E$. Moreover, in the MSN studied, when a user enters the WIT range of the H-AP, it can only download its requested information from the H-AP. If the user enters the WET range of the H-AP, it can simultaneously download the information and harvest energy from the H-AP, since we have $r^E < r^D$. In this MSN, we aim for deploying the $K$ H-APs at the optimal sites in order to maximise the performance of both the WIT and WET.
\subsection{Mobility Model}
\begin{figure}
  \centering

  \includegraphics[width=3.5in]{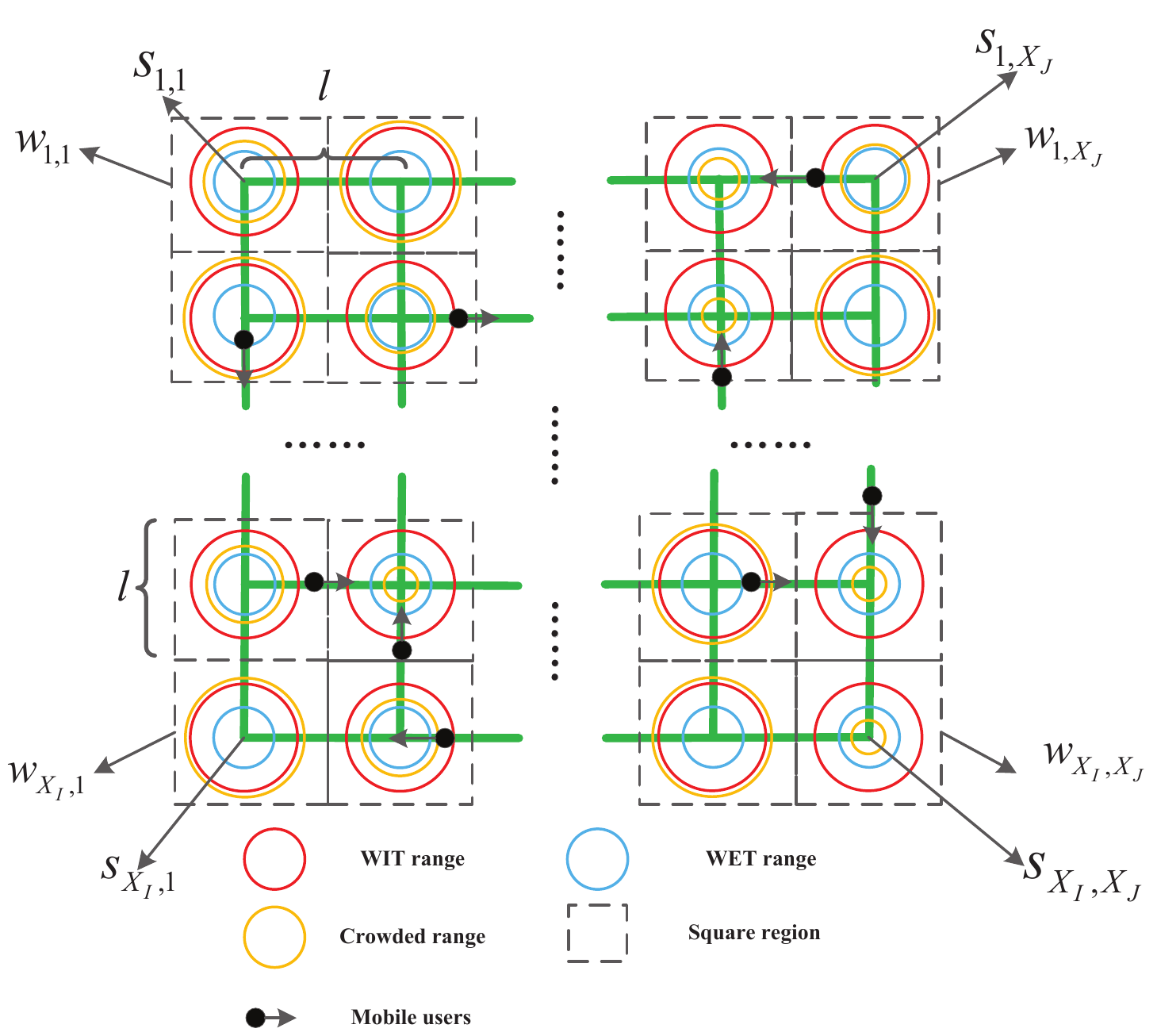}
  \caption{Mobility scenario of city streets}\label{Fig1}
\end{figure}
We assume that all the users move around a grid based city street blocks, by obeying a mobility model more generic than the conventional Manhattan mobility model \cite{2222}. In our mobility model, the users move along the streets. Once they arrive at a crossroad, they may choose any available direction to move forward.  Since users from four different directions may gather at a crossroad, deploying a H-AP at a crossroad is capable of efficiently serving more users. As a result, all the crossroads are regarded as the potential sites for the deployment of the H-APs.

Our city street blocks are illustrated in Fig.\ref{Fig1}, where the green lines indicate the streets in the city, while the red and blue circles represent the WIT ranges and the WET ranges of the H-APs, respectively. The entire area is equally divided into several square regions having the cross roads as their centres, as portrayed in Fig.\ref{Fig1}. We assume that there are $X_I\times X_J=N$ crossroads in the city street blocks, where $X_I$ represents the number of crossroads in the vertical dimension, while $X_J$ represents the number of crossroads in the horizontal dimension. All the crossroads can be denoted by the set $\{s_{i,j} | 1\leq i \leq X_I, 1\leq j \leq X_J\}$, which are also potential sites for the deployment of the H-APs, and users' movements are bounded in the area studied. Generally, a crossroad may have arbitrary number of streets connected to it. However, in our model, we consider that a crossroad is only connected to its nearest peers in our grid based street blocks.  Moreover, a crossroad also has a so-called crowded range for characterising its popularity.  Moving within the crowded range may substantially reduce the speed. A higher crowded range represents that more people gather at this crossroad.  The crowded range of the crossroad $s_{i,j}$ is represented by an orange circle having the radius of $r^C_{i,j}$. The crowded range $r_{i,j}^C$ is an average value according to the statistics of the crossroad $s_{i,j}$. When a user $u_m$ moves outside of the crowded ranges of the crossroads, its speed is assumed to be $v_{m,0}$. If $u_m$ enters the crowded range of the crossroad $s_{i,j}$, it may reduce its speed to $v_{m,i,j}$, which is determined by the user $u_m$ itself and by the popularity of the crossroad $s_{i,j}$.  Without loss of generality, every street connecting a pair of neighbouring crossroads has an identical length of $l$.  Since the short-range communication technique, such as WiFi, bluetooth and etc., is invoked for enabling the D2A communication in the MSN, the WIT range $r^D$ is relatively low. As a result, we may reasonably assume that the street length satisfy the condition of $l>2\max(r^D,r^E,r^C_{i,j})$.

\section{User Mobility Analysis}

In this section, we model the users' mobility pattern by a two-dimensional discrete Markov chain, which may help us in analysing a user's sojourn duration within the WET range of a H-AP and that within its WIT range, respectively.

\subsection{Two-Dimensional Markov Chain Modelling}
As portrayed in Fig.\ref{Fig1}, the entire area is equally divided into a set of square regions having cross roads as their centres. As a result,  a single street is halved into two segments belonging to a pair of neighbouring square regions.  The side length of each square region is $l$, which is equal to the length of a street. The square region having the crossroad $s_{i,j}$ as its centre is denoted by $w_{i,j}$. All the direction choosing probabilities of the user $u_m$ can be denoted by a set of $\mathbb{P}^m=\{p^m_{(i_1,j_1),(i_2,j_2)} | {s_{i_1,j_1}} \text{ and } {s_{i_2,j_2}} \text{ are neighbouring crossroads}\}$. Specifically, $p^m_{(i_1,j_1),(i_2,j_2)}$ represents the probability of the user $u_m$ moving from the crossroad $s_{i_1,j_1}$ to its neighbour $s_{i_2,j_2}$. By defining the user $u_m$ 's stay within the square region $w_{i,j}$ as one of its mobility states, we may model the user $u_m$'s mobility pattern by a two-dimensional Markov chain, which is illustrated in Fig.\ref{markov chain}. Consequently, $p^m_{(i_1,j_1),(i_2,j_2)}$ also represents the probability of $u_m$ transiting from the state $w_{i_1,j_1}$ to the state $w_{i_2,j_2}$.

The stationary probabilities of all the states in the Markov chain of $u_m$ can be denoted by a $1\times X_IX_J$ vector  $\boldsymbol{\phi}^m$, whose element $\phi^m_{(i-1)X_J+j}$ represents the corresponding stationary probability of the state $w_{i,j}$. Similarly, we define a $X_IX_J\times X_IX_J$ matrix of the state transition probabilities $\mathbf{P}^{m'}=\{p^{m'}_{i,j}|1\le i\le X_IX_J, 1\le j\le X_IX_J\}$ by re-arranging the entries in the original set $\mathbb{P}^m$ of the state transition probabilities. The entry $p^m_{(i_1,j_1),(i_2,j_2)}$ of $\mathbb{P}^m$ is then mapped on the entry $p^{m'}_{(i_1-1)X_J+j_1,(i_2-1)X_J+j_2}$ of $\mathbf{P}^{m'}$.
Given that our Markov chain is ergodic and irreducible,  the stationary distribution $\boldsymbol{\phi}^m$ can be readily calculated by
\begin{align}
\begin{cases}
\displaystyle{\boldsymbol{\phi}^m \mathbf{P}^{m'}=\boldsymbol{\phi}^m}\\
\displaystyle{\sum\boldsymbol{\phi}^m=1}
\end{cases}
\end{align}
In the rest of the paper, we denote the stationary probability of the state $w_{i,j}$ by $\phi_{i,j}^m$, which is equal to the element of $\phi^m_{(i-1)X_J+j}$ in the vector $\boldsymbol{\phi}^m$.

\subsection{Total Sojourn Duration of the Square Regions}

\begin{figure}
  \centering

  \includegraphics[width=3.5in]{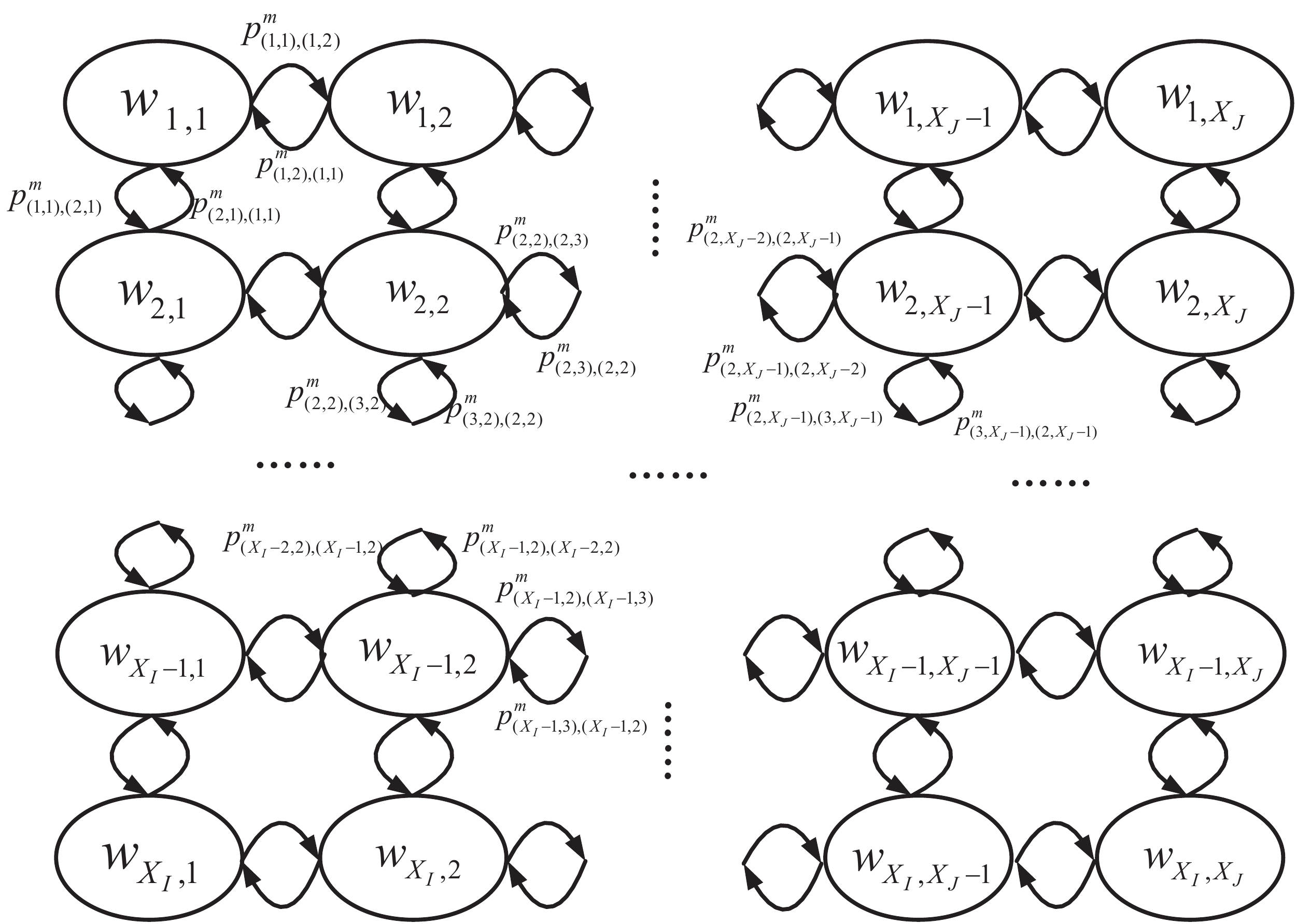}
  \caption{Markov chain model of $u_m$}\label{markov chain}
\end{figure}
Since the users only change their moving directions when they arrive at any of the crossroads, the travel distance of the user $u_m$ within any of the square regions $w_{i,j}$ is $l$. For a specific square region $w_{i,j}$, since we have $r^C_{i,j}<l/2$, the crowded range  of the crossroad $s_{i,j}$ is also included in this square region. As a result, when the user $u_m$ visits the square region $w_{i,j}$, its sojourn duration $D_{m,i,j}^S$ within this region is derived as
\begin{align}\label{eq:DS}
D^S_{m,i,j}=2\left(\frac{r^C_{i,j}}{v_{m,i,j}}+\frac{l/2-r^C_{i,j}}{v_{m,0}}\right)
\end{align}

The sojourn distribution of the user $u_m$ sojourning within a specific square region can be defined as the set $\{\pi^S_{m,1,1},\dots,\pi^S_{m,i,j},\dots,\pi^S_{m,X_I,X_J}\}$.   Note that the probability $\pi^S_{m,i,j}$  is different from the stationary probability $\phi_{i,j}^m$ of the state $w_{i,j}$ in the Markov chain of Fig.\ref{markov chain}, since the derivation of $\phi_{i,j}^m$ only considers the state transition probabilities but the derivation of $\pi^S_{m,i,j}$ further considers the time duration of $u_m$ staying in a specific state. Based on the steady probability $\phi^m_{i,j}$ of the user $u_m$ staying within the square regions $w_{i,j}$ and its corresponding sojourn durations $D^S_{m,i,j}$, the probability of the user $u_m$ sojourning within the square region $w_{i,j}$ can be derived as
\begin{align}\label{eq:piS}
\pi^S_{m,i,j}=\frac{\phi^m_{i,j}D^S_{m,i,j}}{\sum\limits_{i^{'},j^{'}}\phi^m_{i^{'},j^{'}}D^S_{m,i^{'},j^{'}}}
\end{align}

Given a specific observation duration $T$,  the total sojourn duration $\tau^S_{m,i,j}$ of the user $u_m$ in the square region $w_{i,j}$ can be expressed as $\tau^S_{m,i,j}=\pi^S_{m,i,j}T$.

\subsection{Total Sojourn Duration within the WIT and WET ranges}
We firstly aim for obtaining the sojourn duration of the user $u_m$ within the WIT range of the H-AP, if it is deployed at the crossroad $s_{i,j}$. The travel distance of the user $u_m$ within the WIT range should always be $2r^{D}$. If the WIT range $r^D$ is shorter than the crowded range $r^C_{i,j}$, the user $u_m$ always move within the WIT range at a speed of $v_{m,i,j}$. If the WIT range $r^D$ is longer than the crowded range $r^C_{i,j}$, the user $u_m$ may move for a distance of $2r_{i,j}^C$ at a speed of $v_{m,i,j}$ and it may move for another distance of $(2r^D-2r_{i,j}^C)$ at a speed of $v_{m,0}$. As a result, the sojourn duration of the user $u_m$ within the WIT range can be expressed as:
\begin{align}\label{eq:informationrangetime}
D^D_{m,i,j}=
\begin{cases}
\displaystyle{\frac{2r^D}{v_{m,i,j}},\ r^D<r^C_{i,j}}\\
\\
\displaystyle{\frac{2r^C_{i,j}}{v_{m,i,j}}+\frac{2\left(r^D-r^C_{i,j}\right)}{v_{m,0}},\ r^D\ge r^C_{i,j}}
\end{cases}
\end{align}

By sticking to the same methodology, the sojourn duration $D_{m,i,j}^E$ of the user $u_m$ within the WET range is formulated as:
\begin{align}\label{eq:energyrangetime}
D^E_{m,i,j}=
\begin{cases}
\displaystyle{\frac{2r^E}{v_{m,i,j}},\ r^E<r^C_{i,j}}\\
\\
\displaystyle{\frac{2r^C_{i,j}}{v_{m,i,j}}+\frac{2\left(r^E-r^C_{i,j}\right)}{v_{m,0}},\ r^E\ge r^C_{i,j}}
\end{cases}
\end{align}

Furthermore,  the total sojourn duration $\tau_{m,i,j}^{D}$ of the user $u_m$ within the WIT range of the H-AP deployed at the crossroad $s_{i,j}$ during the entire observation duration $T$ is derived as:
\begin{align}\label{eq:tauD}
\tau^D_{m,i,j}=\tau^S_{m,i,j}\frac{D^D_{m,i,j}}{D^S_{m,i,j}}
\end{align}
Similarly,  the total sojourn duration $\tau_{m,i,j}^{E}$ of the user $u_m$ within the WET range of the H-AP deployed at the crossroad $s_{i,j}$ during the entire observation duration $T$ can be derived by substituting $D^E_{m,i,j}$ for $D^D_{m,i,j}$ into \eqref{eq:tauD}.
According to the above analysis, the following remark is obtained:
\begin{remark}
For a specific crossroad $s_{i_0,j_0}$, when the crowded range of other crossroads are fixed, the total sojourn duration of the user $u_m$ in the WIT range of $s_{i_0,j_0}$ is a convex function with respect to $r^C_{i_0,j_0}$. The total sojourn duration $\tau^D_{m,i,j}$ can be maximised when the crowded range $r^C_{i_0,j_0}$ is equal to the WIT range $r^D$.  By contrast, the total sojourn duration of the user $u_m$ within the WIT range of any other crossroad is a monotonously decreasing function with respect to the crowded range $r^C_{i_0,j_0}$ of the crossroad $s_{i_0,j_0}$. Same conclusion can be made on the total sojourn duration of the user $u_m$ within the WET range of the H-AP.
\end{remark}
\begin{IEEEproof}
Please refer to appendix A for the detailed proof.
\end{IEEEproof}

\section{Problem Formulation}
By considering the users' mobility pattern and the popularity of the crossroads, we aim for obtaining an optimal H-AP deployment scheme for maximising either the overall WIT or the overall WET performance of the MSN.
\subsection{WIT Efficiency}
The successful WIT efficiency is defined as the ratio between the actually delivered information amount and the maximum possible delivered information amount of all the users in the MSN studied. We assume that a user may download the information from the H-AP with a constant rate $\mu$, which is readily realised by invoking power control, the WIT efficiency of all the $M$ users during the observation duration $T$ is formulated as
\begin{align}\label{eq:eta}
\eta=\frac{\sum\limits_m\sum\limits_{i,j}\mu \tau^D_{m,i,j}y_{i,j}}{\sum\limits_m\sum\limits_{i,j}\mu \tau^D_{m,i,j}}=\frac{\sum\limits_m\sum\limits_{i,j} \tau^D_{m,i,j}y_{i,j}}{MT}
\end{align}
where $y_{i,j}$ denotes the deployment state  of the H-AP at the crossroad $s_{i,j}$. Furthermore, $y_{i,j}=1$ indicates that an H-AP is deployed at $s_{i,j}$, while $y_{i,j}=0$ indicates that no H-AP is deployed at $s_{i,j}$. All the $y_{i,j}$ form an H-AP deployment matrix $\mathbf{Y}$, having a size of $X_I\times X_J$. The numerator of \eqref{eq:eta} indicates the actually delivered information amount of all the users, given  a specific deployment scheme of the H-APs, while the denominator of \eqref{eq:eta} indicates the theoretically achievable information amount of all the users, if at least an H-AP  is deployed at each crossroad.
According to the above analysis, we have the following remark:
\begin{remark}
If no H-AP is deployed at the crossroad $s_{i_0,j_0}$, the WIT efficiency $\eta$ is a monotonously decreasing function with respect to $r^C_{i_0,j_0}$. By contrast,  if an H-AP is deployed at the corssroad $s_{i_0,j_0}$, the WIT efficiency $\eta$ is a convex function with respect to $r^C_{i_0,j_0}$. Furthermore, the WIT efficiency $\eta$ can be maximised, when we have $r^C_{i_0,j_0} = r^D$.
\end{remark}
\begin{IEEEproof}
Please refer to appendix B for the detailed proof.
\end{IEEEproof}

As a result, in order to improve the WIT efficiency, we should make the crossroads less crowded if no H-AP is deployed, while we should let $r^C_{i,j}=r^D$, if an H-AP is deployed at the crossroad $s_{i,j}$.
\subsection{WET Efficiency}

We define the WET efficiency as the total harvested energy amount by all the users in the network. Due to the  path loss incurred channel attenuation, the power of the received RF signals at the users is determined by the signals' propagation distance. For example, when the user $u_m$ enters the WET range of the H-AP deployed at the crossroad $s_{i,j}$, the distance from $u_m$ to the H-AP firstly reduce. Since the user keeps moving, the distance between $u_m$ and the H-AP at $s_{i,j}$ will first decrease from $r^E$ to 0 and increase to $r^E$ again. According to \cite{2121}, the path-loss between the H-AP and $u_m$, when they are $\varsigma_m$ metres away from each other, can be expressed as
\begin{align}\label{eq:channel}
h_m=G_tG_r\left(\frac{c}{4\pi f\varsigma_0}\right)^2\left(\frac{\varsigma_0}{\varsigma_m}\right)^\beta,\ \varsigma_m>\varsigma_0
\end{align}
where $G_t$ and $G_r$ are the transmit and receive antenna gains, respectively, while $c$ is the speed of the light, $f$ is the carrier frequency, and $\beta$ is the path loss exponent. Furthermore, $\varsigma_0$ and $\varsigma_m$ are the reference distance and the distance between $u_m$ and the H-AP, respectively. Denoting the path loss at the distance of $\varsigma_0$ as $h_0$, the path-loss model of \eqref{eq:channel} can be re-formulated as
\begin{align}
h_m=h_0\left(\frac{\varsigma_0}{\varsigma_m}\right)^\beta,\ \varsigma_m>\varsigma_0
\end{align}
In the near-field of the transmitters, say  $\varsigma_m\le \varsigma_0$, we assume that the path-loss is  $h_m=h_0$. For the short-range communication,  $\varsigma_0$ is normally a single metre \cite{2121}, which is even lower than the minimum among the WIT range $r_D$, the WET range $r^E$ and the crowded range $r^C_{i,j}$. As a result, the energy harvested by the user $u_m$, when it enters the transmission range of the H-AP deployed at the crossroad $s_{i,j}$, can be formulated as
\begin{align}\label{eq:varepsilon}
\varepsilon_{m,i,j}=
\begin{cases}
\displaystyle{2\delta\left(\int_{\varsigma_0}^{r^E}\frac{Ph_m}{v_{m,i,j}}d{\left(\varsigma_m\right)}+\frac{Ph_0\varsigma_0}{v_{m,i,j}}\right),\ r^E\le r^C_{i,j}}\\ \\
\displaystyle{2\delta\left(\int_{\varsigma_0}^{r^C_{i,j}}\frac{Ph_m}{v_{m,i,j}}d{\left(\varsigma_m\right)}+\int_{r^C_{i,j}}^{r^E}\frac{Ph_m}{v_{m,0}}d{\left(\varsigma_m\right)}\right.}\\ \\
\displaystyle{\ \ \ \ \ \ \ \ \left.+\frac{Ph_0\varsigma_0}{v_{m,i,j}}\right),\ r^E>r^C_{i,j}}
\end{cases}
\end{align}
where $P$ indicates the transmitting power of the H-AP, while $\delta$ indicates the efficiency of rectifying the received RF signals to the direct current (DC) and we make $\delta=80\%$ without loss of generality. Finally, the closed-form equation of the energy harvested by the user $u_m$ can be derived from \eqref{eq:varepsilon} as
\begin{align}\label{eq:varepsilon2}
\varepsilon_{m,i,j}=
\begin{cases}
\displaystyle{2\delta\frac{Ph_0\varsigma_0}{(1-\beta)v_{m,i,j}}\left(\left(\frac{r^E}{\varsigma_0}\right)^{1-\beta}-\beta\right),\ r^E\le r^C_{i,j}}\\ \\
\displaystyle{2\delta\frac{Ph_0\varsigma_0}{(1-\beta)}\left(\left(\frac{1}{v_{m,i,j}}-\frac{1}{v_{m,0}}\right)\left(\frac{r^C_{i,j}}{\varsigma_0}\right)^{1-\beta}\right.}\\ \\
\displaystyle{\ \ \ \ \ \ \ \ \left.+\frac{1}{v_{m,0}}\left(\frac{r^E}{\varsigma_0}\right)^{1-\beta}-\frac{\beta}{v_{m,i,j}}\right),\ r^E>r^C_{i,j}}
\end{cases}
\end{align}
As shown in \eqref{eq:varepsilon2}, the amount $\varepsilon_{m,i,j}$ of energy harvested by the user $u_m$ at the crossroad $s_{i,j}$ increases, as the crowded range $r_{i,j}^C$ of $s_{i,j}$ increases. If we keep increasing $r_{i,j}^C$ beyond $r_{i,j}^E$, the energy $\varepsilon_{m,i,j}$ converges to a constant value.

Given the total observation duration $T$, the sojourn duration of the user $u_m$ within the WET range of the H-AP deployed at the crossroad $s_{i,j}$ is derived as $\tau_{m,i,j}^E$. As a result, the number of times that $u_m$ moves within the WET range  of the H-AP deployed at the crossroad $s_{i,j}$ during the entire observation duration $T$ can be calculated as $\frac{\tau^E_{m,i,j}}{D^E_{m,i,j}}$. Therefore,  the total energy  harvested by the user $u_m$ during the total observation duration $T$ can be formulated as
\begin{align}\label{eq:12}
\varepsilon_m=\sum_{i,j}\frac{\tau^E_{m,i,j}}{D^E_{m,i,j}}\varepsilon_{m,i,j}y_{i,j}
\end{align}

Since the energy  harvested by the users have to be stored in their batteries, energy harvesting may not operate, once their batteries are fully charged. Without loss of generality, we assume that all the users' battery capacity are $Q$. Hence, the actual energy  harvested by the user $u_m$ can be expressed as $\min(\varepsilon_m,Q)$. Furthermore, the WET efficiency can be obtained as
\begin{align}\label{eq:totalvarepsilon}
\varepsilon=\sum_m\min\left(\varepsilon_m,Q\right)
\end{align}

According to the above analysis,  we have the remark as follows:
\begin{remark}
When the crowded range of the crossroad $s_{i_0,j_0}$ satisfies  the condition of $r^C_{i_0,j_0}>r^E$, the WET efficiency does not increase as the crowded range $r^C_{i_0,j_0}$ increases, regardless of whether an H-AP is deployed at the crossroad $s_{i_0,j_0}$ or not.
\end{remark}
\begin{IEEEproof}
Please refer to appendix C for the detailed proof.
\end{IEEEproof}

Remark 3 tells us that in order to improve the WET efficiency, we should make the crowded range $r^C_{i_0,j_0}$ of the crossroad $s_{i_0,j_0}$ smaller, when it satisfies the condition $r^C_{i_0,j_0}>r^E$.

\subsection{H-AP Deployment Schemes}

We will study three deployment schemes of the H-APs for serving different purposes, namely the balanced deployment scheme (B-deployment), the energy harvesting oriented deployment scheme (E-deployment) and the information downloading oriented deployment scheme (I-deployment).

In order to maximise  the WIT efficiency $\eta$ by ensuring that the total energy harvested by the users is higher than a threshold, we propose the B-deployment scheme, which can be formulated as the following optimisation problem:
\begin{align}
&\text{Objective:}&\max_{\mathbf{Y}}\ \eta,\label{B-deployment}\\
&\text{Subject to}:&\sum_{i,j}y_{i,j}=K,\tag{14a}\label{constraint-14a}\\
& &y_{i,j}=0\ or\ 1,\tag{14b}\label{constraint-14b}\\
& &\varepsilon\ge \alpha\varepsilon_{\max},\tag{14c}\label{constraint-14c}
\end{align}
where  the parameter $\alpha \in [0,1]$ is introduced for adjusting the designer's preference on the WIT performance and the WET performance. A higher value of $\alpha$ indicates that the designer is inclined to obtain a higher WET performance, while a lower value of $\alpha$ indicates that the design is inclined to the WIT performance. The parameter $\varepsilon_{\max}$ in \eqref{constraint-14c} is the maximum energy that can be harvested by the users, which can be obtained by solving the following optimisation problem:
\begin{align}
&\text{Objective:}&\max_{\mathbf{Y}}\ \varepsilon,\label{E-deployment}\\
&\text{Subject to:}&\sum_{i,j}y_{i,j}=K,\tag{15a}\label{constraint-15a}\\
& &y_{i,j}=0\ or\ 1.\tag{15b}\label{constraint-15b}
\end{align}

The optimisation problem \eqref{E-deployment} is regarded as the E-deployment scheme. Note that if we have $\alpha=1$ in \eqref{constraint-14c}, the optimisation problem \eqref{B-deployment} is degenerated to \eqref{E-deployment}. If we have $\alpha=0$ in \eqref{constraint-14c}, the B-deployment scheme is degenerated to the I-deployment scheme, which can be formulated as
\begin{align}
&\text{Objective:}&\max_{\mathbf{Y}}\ \eta,\label{I-deployment}\\
&\text{Subject to:}&\sum_{i,j}y_{i,j}=K,\tag{16a}\label{constraint-16a}\\
& &y_{i,j}=0\ or\ 1.\tag{16b}\label{constraint-16b}
\end{align}

\section{Solutions of the H-AP Deployment}
\subsection{Exhaustive Searching Method}
Since the deployment solution $\mathbf{Y}$ is an integer matrix, the above three problems cannot be solved in a polynomial time. The exhaustive search is capable of traversing all the possible deployment solutions and it thus choose the optimal one. This is the most inefficient method since its complexity is as high as $\mathbf{C}_{X_IX_J}^K$. However, when the solution space is small, exhaustive search can be adopted for finding the optimal solution with the tolerant complexity.  Let us define $g(\mathbf{Y})$ as the generic objective function with respect to the H-AP's deployment indicator matrix  $\mathbf{Y}$. Hence, we have $g(\mathbf{Y}) = \eta$ for both the I-deployment problem and the B-Deployment scheme. We also have $g(\mathbf{Y}) = \varepsilon$ for the E-deployment problem. The pseudo code of the exhaustive search is provided in  Algorithm \ref{alg:alg1}.
\begin{algorithm}[!t]
\caption{exhaustive searching method}
\label{alg:alg1}
\footnotesize
\begin{algorithmic}[1]
\STATE Define all the $\mathbf{C}_{X_IX_J}^K$ possible H-AP deployment set as $\Psi=\{\mathbf{Y}_1,\mathbf{Y}_2,\dots,\mathbf{Y}_{\mathbf{C}_{X_IX_J}^K}\}$, in which each element in $\Psi$ is an optional deployment. Define $\mathbf{Y}^*=\mathbf{Y}_1$, $g(\mathbf{Y}^*)=0$. Initialize deployment subscript $p=1$;
\WHILE {$p\le\mathbf{C}_{X_IX_J}^K$}
	\STATE Calculate $g(\mathbf{Y}_p)$;
	\STATE If $g(\mathbf{Y}_p)\ge g(\mathbf{Y}^*)$ and $\mathbf{Y}_p$ satisfies all the constraints, let $\mathbf{Y}^*=\mathbf{Y}_p$ and $g(\mathbf{Y}^*)=g(\mathbf{Y}_p)$;
	\STATE $p=p+1$;
\ENDWHILE
\RETURN The solution of H-AP deployment $\mathbf{Y}^*$.
\end{algorithmic}
\end{algorithm}

Since the exhaustive search is an aimless searching method, we have to traverse all the possible deployment  solution before finding the optimal one. Since the objective function of the I-deployment problem \eqref{I-deployment} is linear, a directional searching method can be adopted for reducing the complexity. By letting $k_{i,j}=\sum_m\tau^D_{m,i,j}/MT$, the WIT efficiency $\eta$ expressed in \eqref{eq:eta} can be re-formulated as
\begin{align}\label{eq:eta2}
\eta=\sum\limits_{i,j}k_{i,j}y_{i,j}
\end{align}
 In order to maximize the WIT efficiency $\eta$, we should deploy an H-AP at the crossroad $s_{i,j}$ associated with a higher coefficient $k_{i,j}$. The pseudo code of the directional searching is provided in  Algorithm \ref{alg:alg2}.
\begin{algorithm}[!t]
\caption{directional searching method}
\label{alg:alg2}
\footnotesize
\begin{algorithmic}[1]
\STATE Calculate all the value of $k_{i,j}$ in \eqref{eq:eta2}. Initialize deployment subscript $p=1$, and deployment matrix $\mathbf{Y}$;
\WHILE {$p\le K$}
	\STATE Find the max $k_{i,j}$;
	\STATE Let $y_{i,j}=1$;
	\STATE Let $k_{i,j}=0$;
\ENDWHILE
\RETURN The solution of H-AP deployment $\mathbf{Y}$.
\end{algorithmic}
\end{algorithm}

By adopting the directional searching, the complexity can be reduced to $\mathcal{O}(K)$. By contrast,  since the objective functions in the E-deployment  problem is non-linear, the directional searching cannot be adopted for finding the optimal deployment solution. Furthermore, since there is a non-linear constraint on the energy harvesting performance, the directional searching cannot be adopted for finding the optimal solution of the B-deployment problem.
\subsection{H-AP Deploying Probability}

Observe from \eqref{B-deployment}-\eqref{I-deployment} that the optimisation problems of the B-deployment scheme, the E-deployment scheme and the I-deployment scheme all belong to the integer programming, since the deployment indicator $y_{i,j}$ at the crossroad $s_{i,j}$ is either zero or one, which cannot be solved in polynomial time. As a result, we transform these three integer programming problem into the real number based optimisation problem by relaxing the original integer constraint on $y_{i,j}$ to the constraint that $y_{i,j}$ is a real number within the region $[0,1]$. This relaxation is reasonable since a real valued $y_{i,j}$ can be regarded as the probability of an H-AP being deployed at the crossroad $s_{i,j}$.

We will elaborate on  how the constraint relaxation operates in solving the optimisation problem of the B-deployment as an example, since both the E-deployment scheme and the I-deployment scheme are special cases of the B-deployment scheme. We introduce a new variable $R_m= \min(\varepsilon_m, Q)$ for representing the actual energy harvested by the user $u_m$. The constraint \eqref{constraint-14c} can be then rewritten as
\begin{align}
\begin{cases}
R_m\le\varepsilon_m\notag\\
R_m\le Q,\notag\\
\sum\limits_mR_m\ge\alpha\varepsilon_{\max}.\notag
\end{cases}
\end{align}
Accordingly, the original optimisation problem of \eqref{B-deployment} can be reformulated as
\begin{align}\label{B-deployment-real}
&\text{Objective:}&\max_{\mathbf{Y},R_m}\ \sum\limits_{i,j}k_{i,j}y_{i,j},\notag\\
&\text{Subject to:}&\sum_{i,j}y_{i,j}=K,\notag\\
& &0\le y_{i,j}\le 1,\notag\\
& &R_m\le\varepsilon_m,\notag\\
& &R_m\le Q,\notag\\
& &\sum\limits_mR_m\ge\alpha\varepsilon_{max},
\end{align}
where $k_{i,j}$ is defined in Section V-A. Since both the objective function and the constraints are linear with respect to $y_{i,j}$ and $R_m$, the optimisation problem \eqref{B-deployment-real} can be solved by adopting the simplex method \cite{2323}.

Similarly, the reformulated optimisation problem of the E-deployment scheme can be expressed as
\begin{align}\label{E-deployment-real}
&\text{Objective:}&\max_{\mathbf{Y},R_m}\ \sum\limits_{m}R_m,\notag\\
&\text{Subject to:}&\sum_{i,j}y_{i,j}=K,\notag\\
& &0\le y_{i,j}\le 1,\notag\\
& &R_m\le\varepsilon_m,\notag\\
& &R_m\le Q,
\end{align}
and the reformulated optimisation problem of the I-deployment scheme can be expressed as
\begin{align}\label{I-deployment-real}
&\text{Objective:}&\max_{\mathbf{Y}}\ \sum\limits_{i,j}k_{i,j}y_{i,j},\notag\\
&\text{Subject to:}&\sum_{i,j}y_{i,j}=K,\notag\\
& &0\le y_{i,j}\le 1.
\end{align}
Both these two optimisation problems can also be solved by adopting the simplex method.

\subsection{Branch-and-bound approach}

Since the exhaustive searching cannot solve the optimisation problem of the H-AP's deployment in a polynomial time and the directional searching is only capable of solving the optimisation problem of the I-deployment, we need a generic low-complexity algorithm for finding the optimal solution of all the three deployment scheme.  As a result, the branch-and-bound approach is adopted for solving these optimisation problems by sacrificing the optimality to some extent. The key steps of the branch-and-bound approach are summarised as follows, where the B-deployment optimisation problem is chosen as an example:

\textbf{STEP 1:} Denote the upper bound and the lower bound of the H-AP's deployment indicator $y_{i,j}$ at the crossroad $s_{i,j}$ as $\Delta^U_{i,j}$ and $\Delta^D_{i,j}$, respectively. Let us initialise $\Delta^U_{i,j}=1$ and $\Delta^D_{i,j}=0$ for each $y_{i,j}$. The set of all $\Delta^U_{i,j}$ is denoted as $\boldsymbol{\Delta}^U$ , while the set of all $\Delta^D_{i,j}$ is denoted as $\boldsymbol{\Delta}^D$.

\textbf{STEP 2:} We then transform original integer  programming problem into the following real valued optimisation problem as
\begin{align}\label{eq:real-problem}
&\text{Objective:}&\max_{\mathbf{Y},R_m}\ g(\mathbf{Y}),\notag\\
&\text{Subject to:}&\Delta^D_{i,j}\le y_{i,j}\le \Delta^U_{i,j},\forall s_{i,j},\notag\\
& &\sum_{i,j}y_{i,j}=K, \eqref{constraint-14c},
\end{align}
where $g(\mathbf{Y})=\sum\limits_{i,j}k_{i,j}y_{i,j}$.
We can use the same method for solving the real programming problem in Section V-B, while the optimal solution of all the $y_{i,j}$ of \eqref{eq:real-problem} is denoted as the set $\mathbf{Y}^*$.

\textbf{STEP 3:} If the resultant in $\mathbf{Y}^*$ is an integer matrix, $\mathbf{Y}^*$ is the final solution. Otherwise, let $\boldsymbol{\Delta}^{U1}=\boldsymbol{\Delta}^{U2}=\boldsymbol{\Delta}^U$ and $\boldsymbol{\Delta}^{D1}=\boldsymbol{\Delta}^{D2}=\boldsymbol{\Delta}^D$. Then, arbitrarily choose a non-integer element $y_{i,j}$ in $\mathbf{Y}^*$. Let $\Delta^{U1}_{i,j}=\lfloor y_{i,j}\rfloor$ and $\Delta^{D2}_{i,j}=\lceil y_{i,j}\rceil$, where $\lfloor x\rfloor$ and $\lceil x\rceil$ indicates the largest integer smaller than $x$ and the smallest integer larger than $x$, respectively.

\textbf{STEP 4:} Divide the optimisation problem \eqref{eq:real-problem} into the following two sub-problems
\begin{align}\label{eq:real-problem1}
&\text{Objective:}&\max_{\mathbf{Y},R_m}\ g(\mathbf{Y}),\notag\\
&\text{Subject to:}&\Delta^{D1}_{i,j}\le y_{i,j}\le \Delta^{U1}_{i,j},\forall s_{i,j},\notag\\
& &\sum_{i,j}y_{i,j}=K, \eqref{constraint-14c},
\end{align}
and
\begin{align}\label{eq:real-problem2}
&\text{Objective:}&\max_{\mathbf{Y},R_m}\ g(\mathbf{Y}),\notag\\
&\text{Subject to:}&\Delta^{D2}_{i,j}\le y_{i,j}\le \Delta^{U2}_{i,j},\forall s_{i,j},\notag\\
& &\sum_{i,j}y_{i,j}=K, \eqref{constraint-14c},
\end{align}

Both of these two sub-problems can be solved by exploiting the simplex method of Section V-B. The optimal solutions of these two sub-problems are denoted as $\mathbf{Y}^*_1$ and $\mathbf{Y}^*_2$, respectively. If we have $g(\mathbf{Y}^*_1)>g(\mathbf{Y}^*_2)$, we have $\boldsymbol{\Delta}^U=\boldsymbol{\Delta}^{U1}$, $\boldsymbol{\Delta}^D=\boldsymbol{\Delta}^{D1}$, and $\mathbf{Y}^*=\mathbf{Y}^*_1$. Otherwise, we have $\boldsymbol{\Delta}^U=\boldsymbol{\Delta}^{U2}$, $\boldsymbol{\Delta}^D=\boldsymbol{\Delta}^{D2}$ and $\mathbf{Y}^*=\mathbf{Y}^*_2$.

The pseudo code of our branch-and-bound approach  is provided in  Algorithm \ref{alg:alg3}. The complexity of the branch-and-bound approach is $\mathcal{O}(2K)$, which is much lower than that of the exhaustive searching method.

\begin{algorithm}[!t]
\caption{branch-and-bound algorithm to obtain the solution of H-AP deployment.}
\label{alg:alg3}
\footnotesize
\begin{algorithmic}[1]
\STATE Define the upper and lower constraint bound of the optimisation variable as $\boldsymbol{\Delta}^U$ and $\boldsymbol{\Delta}^D$ respectively. Initialize the value of each element in these two constraint sets, as stated at STEP 1;
\STATE Solve the transformed real optimisation problem \eqref{eq:real-problem} and obtain the H-AP deployment $\mathbf{Y}^*$, as stated at STEP 2;
\WHILE {At least one element $y_{i,j}$ in $\mathbf{Y}^*$ is not an integer}
	\STATE Separate constraints $\boldsymbol{\Delta}^U$ and $\boldsymbol{\Delta}^D$ into two divided parts, as stated at STEP 3;
	\STATE Solve two sub-problem \eqref{eq:real-problem1} and \eqref{eq:real-problem2}, and obtain the corresponding H-AP deployment $\mathbf{Y}^*_1$ and $\mathbf{Y}^*_2$;
	\STATE Compare two objective value with $\mathbf{Y}^*_1$ and $\mathbf{Y}^*_2$. If $g(\mathbf{Y}^*_1)>g(\mathbf{Y}^*_2)$, let $\boldsymbol{\Delta}^U=\boldsymbol{\Delta}^{U1}$, $\boldsymbol{\Delta}^D=\boldsymbol{\Delta}^{D1}$, and $\mathbf{Y}^*=\mathbf{Y}^*_1$. Else let $\boldsymbol{\Delta}^U=\boldsymbol{\Delta}^{U2}$, $\boldsymbol{\Delta}^D=\boldsymbol{\Delta}^{D2}$ and $\mathbf{Y}^*=\mathbf{Y}^*_2$, as stated at STEP 4;
\ENDWHILE
\RETURN The solution of H-AP deployment $\mathbf{Y}^*$.
\end{algorithmic}
\end{algorithm}

\section{Performance Evaluation}
\begin{table}
\newcommand{\tabincell}[2]{\begin{tabular}{@{}#1@{}}#2\end{tabular}}
\centering
\small
\caption{The description of the main mathematical symbols}
\label{tab:1}
\scalebox{0.9}{
\begin{tabular}{|l|l|}
\hline
\textbf{Symbols}     &\textbf{The meaning of these symbols} \\
\hline
$M$ & number of users\\
\hline
$N$ & number of candidate sites to deploy H-APs\\
\hline
$K$ & number of H-APs to be deployed\\
\hline
$u_m$ & the $m$-th user, $1\le m\le M$\\
\hline
$s_n$ & the $n$-th candidate site to deploy H-AP\\
\hline
$X_I$,$X_J$ & number of crossroads in vertical and horizontal dimension\\
\hline
$s_{i,j}$ & the crossroad with the coordinate of $[i,j]$\\
\hline
$w_{i,j}$ & the square region of the crossroad with the coordinate of $[i,j]$\\
\hline
$r^D$ & WIT range of the H-AP\\
\hline
$r^E$ & WET range of the H-AP\\
\hline
$r^C_{i,j}$ & crowded range of the crossroad with the coordinate of $[i,j]$\\
\hline
$\mathbf{y}$ & H-AP deployment vector\\
\hline
$\phi^m_{i,j}$ & stationary probability of $u_m$ staying in $w_{i,j}$\\
\hline
$\pi^m_{i,j}$ & stationary probability of $u_m$ staying in $w_{i,j}$ in any time\\
\hline
$D^S_{m,i,j}$ & sojourn duration of $u_m$ visiting $w_{i,j}$\\
\hline
$D^D_{m,i,j}$ & sojourn duration of $u_m$ visiting the WIT range of $s_{i,j}$\\
\hline
$D^E_{m,i,j}$ & sojourn duration of $u_m$ visiting the WET range of $s_{i,j}$\\
\hline
$T$ & observation duration\\
\hline
$\tau^S_{m,i,j}$ &total sojourn duration of $u_m$ staying in $w_{i,j}$\\
\hline
$\tau^D_{m,i,j}$ &total sojourn duration of $u_m$ staying in the WIT range of $s_{i,j}$\\
 \hline
$\tau^E_{m,i,j}$ &total sojourn duration of $u_m$ staying in the WET range of $s_{i,j}$\\
 \hline
$\eta$ & WIT efficiency\\
 \hline
$\varepsilon_{m,i,j}$ & energy harvesting amount of $u_m$ visiting $w_{i,j}$ at each time\\
\hline
$\varepsilon$ & WET efficiency\\
\hline
\end{tabular}}
\end{table}
In this section, numerical results are provided for evaluating the performance of the three deployment schemes of the H-AP. The users in the MSN studied move within the city street blocks having $5\times 5$ crossroads in our simulation and each crossroad has a different crowded range. The moving speed of the users in different spaces are arbitrary. However, the speed of a user moving within a crowded range is lower than the speed moving outside of a crowded range. Furthermore, the probability of a user changing its moving direction at a crossroad is also arbitrarily chosen. The parameter settings are summarised in TABLE.\ref{tab:2}, if there is no specific statement.
\begin{table}[!t]
\centering
\caption{Parameter settings}
\label{tab:2}
\begin{tabular}{|c|c|}
\hline
Number of users $M$ & 100  \\
\hline
Total simulation time $T$ & 10 hours  \\
\hline
Number of H-APs $K$ & 8  \\
\hline
Street length $l$ & 200m  \\
\hline
WIT range $r^D$ & 50m  \\
\hline
WET range $r^E$ & 10m  \\
\hline
Crowd range $r^C_{i,j}$ & 5m-60m  \\
\hline
Reference distance $\lambda_0$  & 1m  \\
\hline
Path loss at $\lambda_0$  & 0.003  \\
\hline
Battery capacity $Q$  & 1J  \\
\hline
Importance parameter $\alpha$  & 0.97  \\
\hline
Energy transmitting power $P$  & 1W  \\
\hline
\end{tabular}
\end{table}

\subsection{Accuracy of Markov analysis}
\begin{figure}[!t]
  \centering
  \includegraphics[width=3.5in]{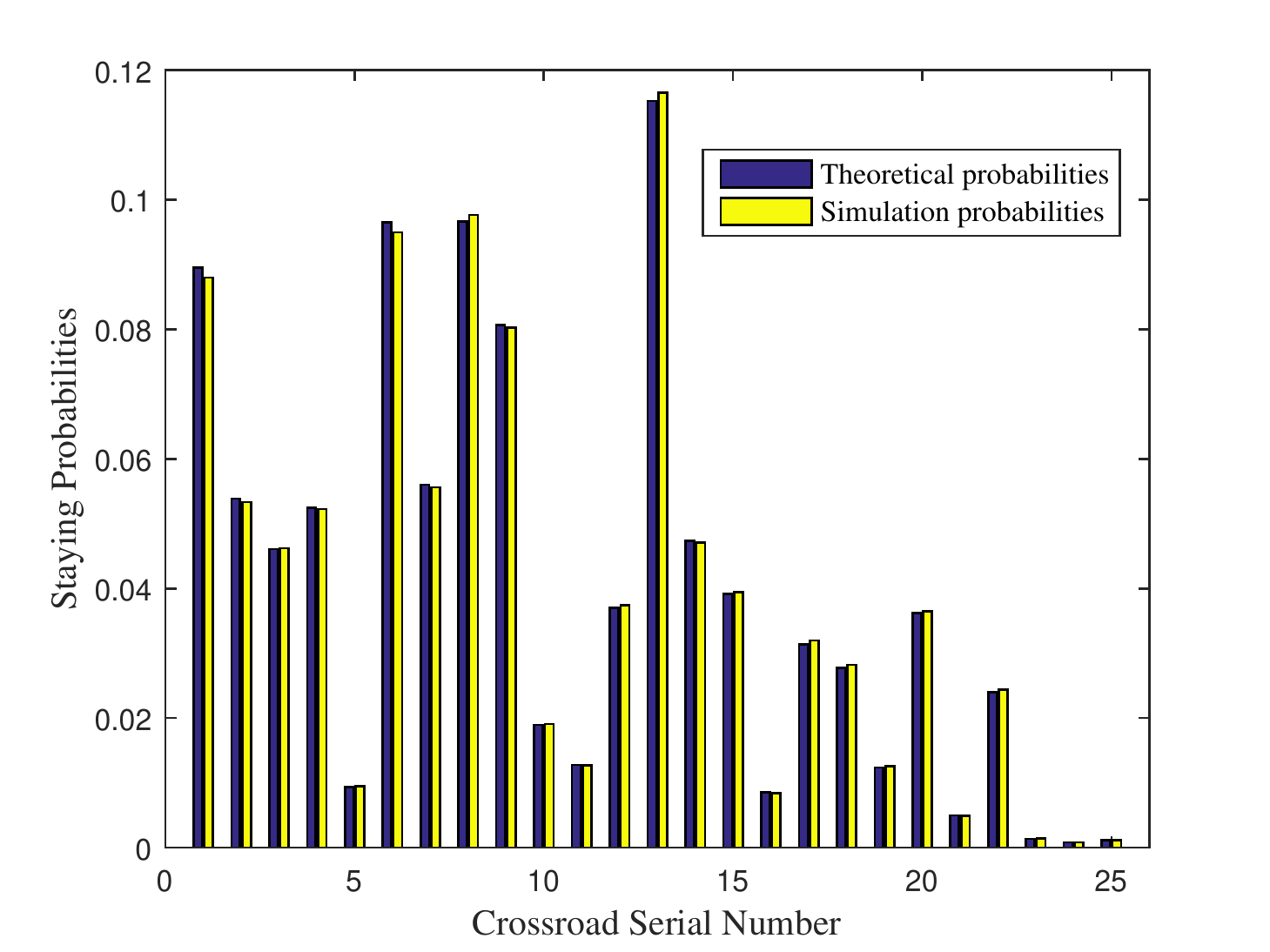}
  \caption{Stationary probabilities of a user staying in different crossroads}\label{confirm_markov}
\end{figure}
 We evaluate a single user's stationary probability of staying at different crossroads by both the Markov chain based analysis and by the Monte-Carlo simulation. Observe from Fig.\ref{confirm_markov} that our analytical results perfectly match the simulation results, which demonstrates the accuracy of our Markov chain aided analysis.

\subsection{Exhaustive Searching v.s. Branch-and-Bound}
\begin{figure}[!t]
  \centering
  \includegraphics[width=4in]{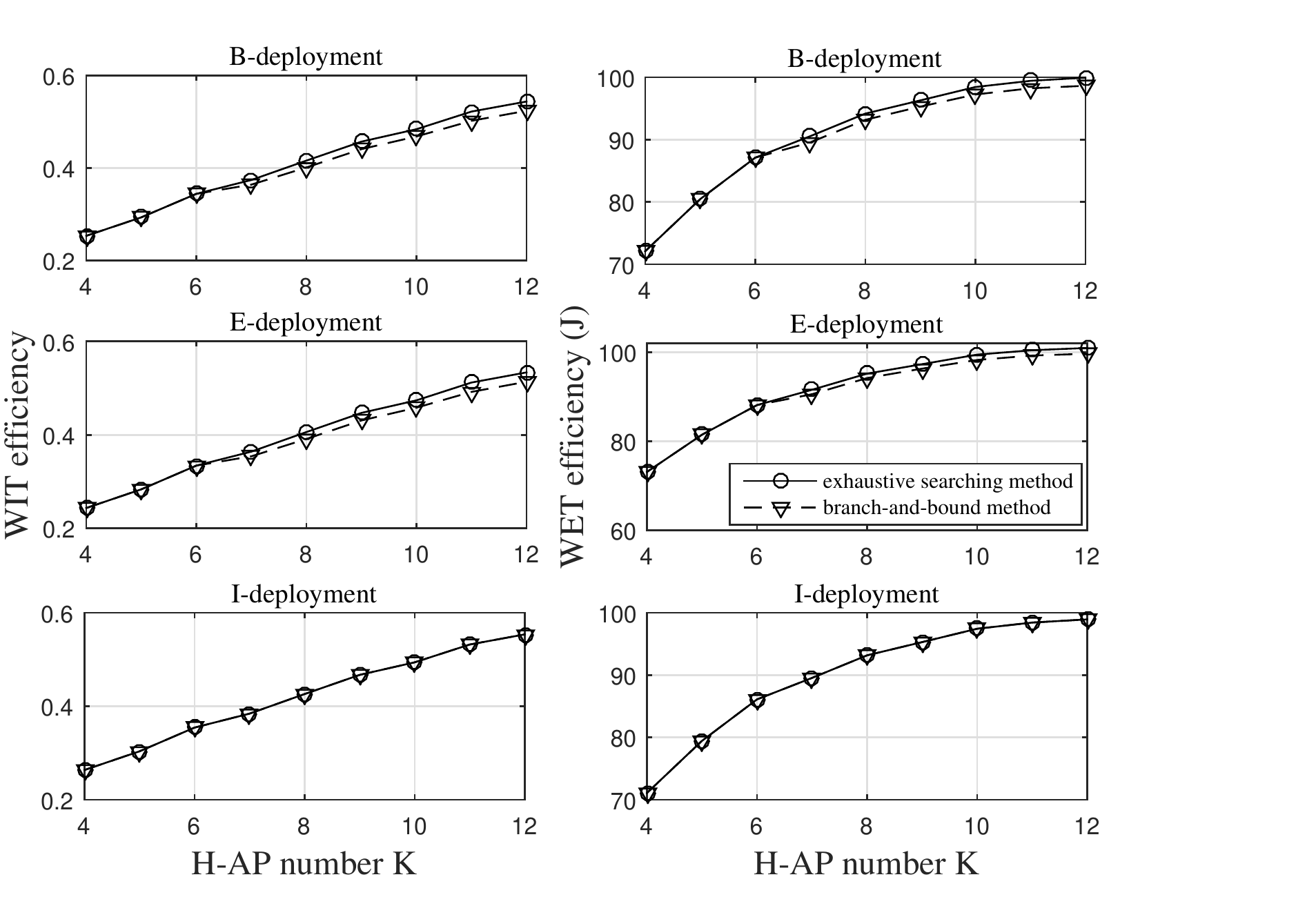}
  \caption{Solutions between two methods}\label{method_com}
\end{figure}
We firstly compare the exhaustive searching method to the branch-and-bound counterpart in terms of the WIT and the WET efficiency. Specifically, the exhaustive searching is capable of finding the optimal deployment solution, while the branch-and-bound based method may only find the sub-optimal deployment solution. Observe from Fig.\ref{method_com} that the solutions obtained by relying on the branch-and-bound approach are always worse  than those of exhaustive searching method, since the branch-and-bound approach sacrifice the optimality for a lower complexity. Note that in the I-deployment scheme, the solutions of these two methods perform the same. This is because the first iteration of the branch-and-bound approach is capable of producing an integer deployment solution $\mathbf{Y}$.  Moreover, observe from Fig.\ref{method_com} that the performance gaps between the solutions of the branch-and-bound method and those of the exhaustive searching method  are negligible.  Hence, when the solution space is very large, the branch-and-bound approach can be adopted for reducing the computational complexity  without remarkable performance degradation.

\subsection{Popularity of Crossroads}

\begin{figure}[!t]
  \centering
  \includegraphics[width=3in]{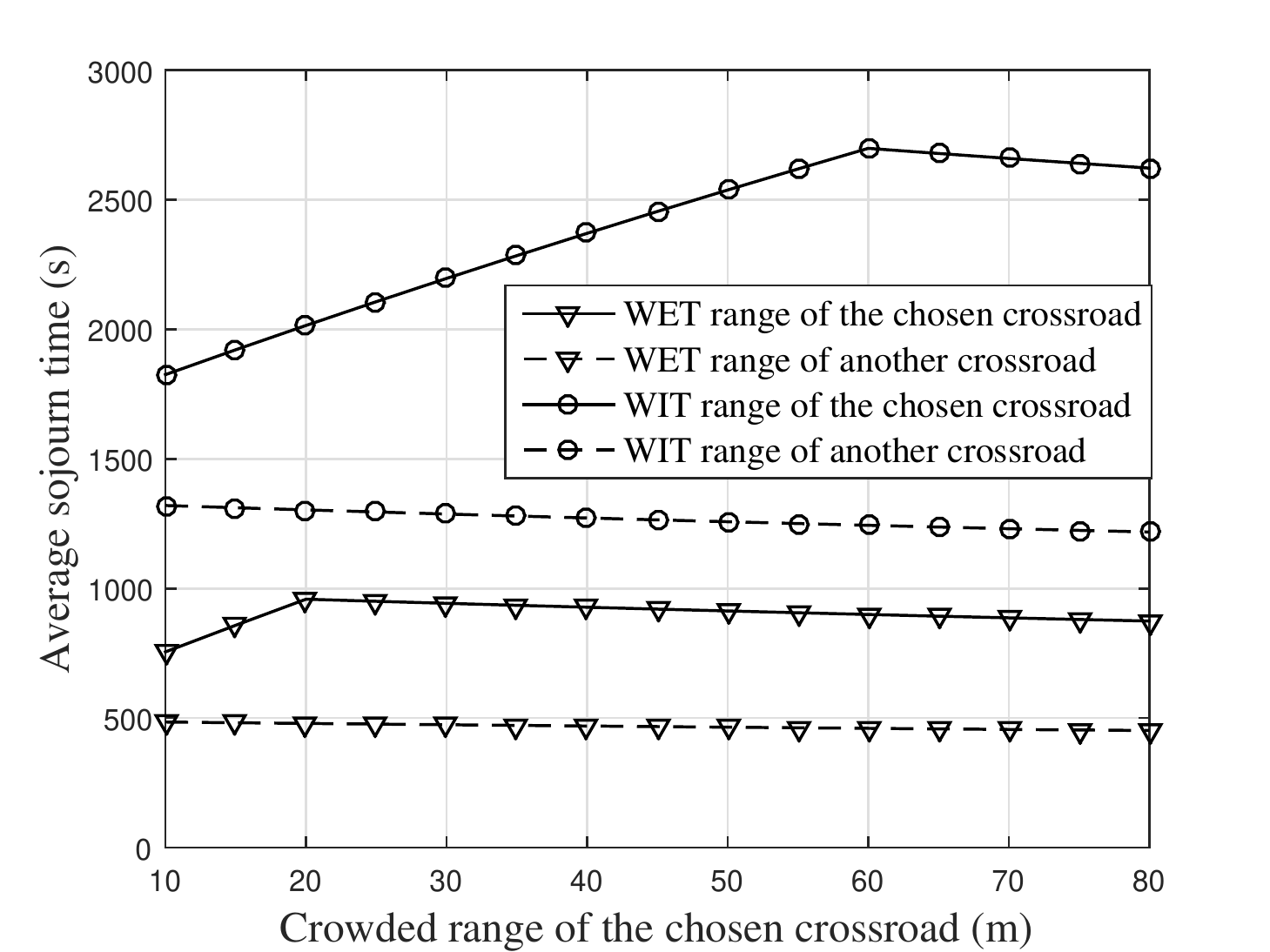}
  \caption{Sojourn time of users at crossroads versus crowded ranges}\label{rC}
\end{figure}
\begin{figure}[!t]
  \centering
  \includegraphics[width=3in]{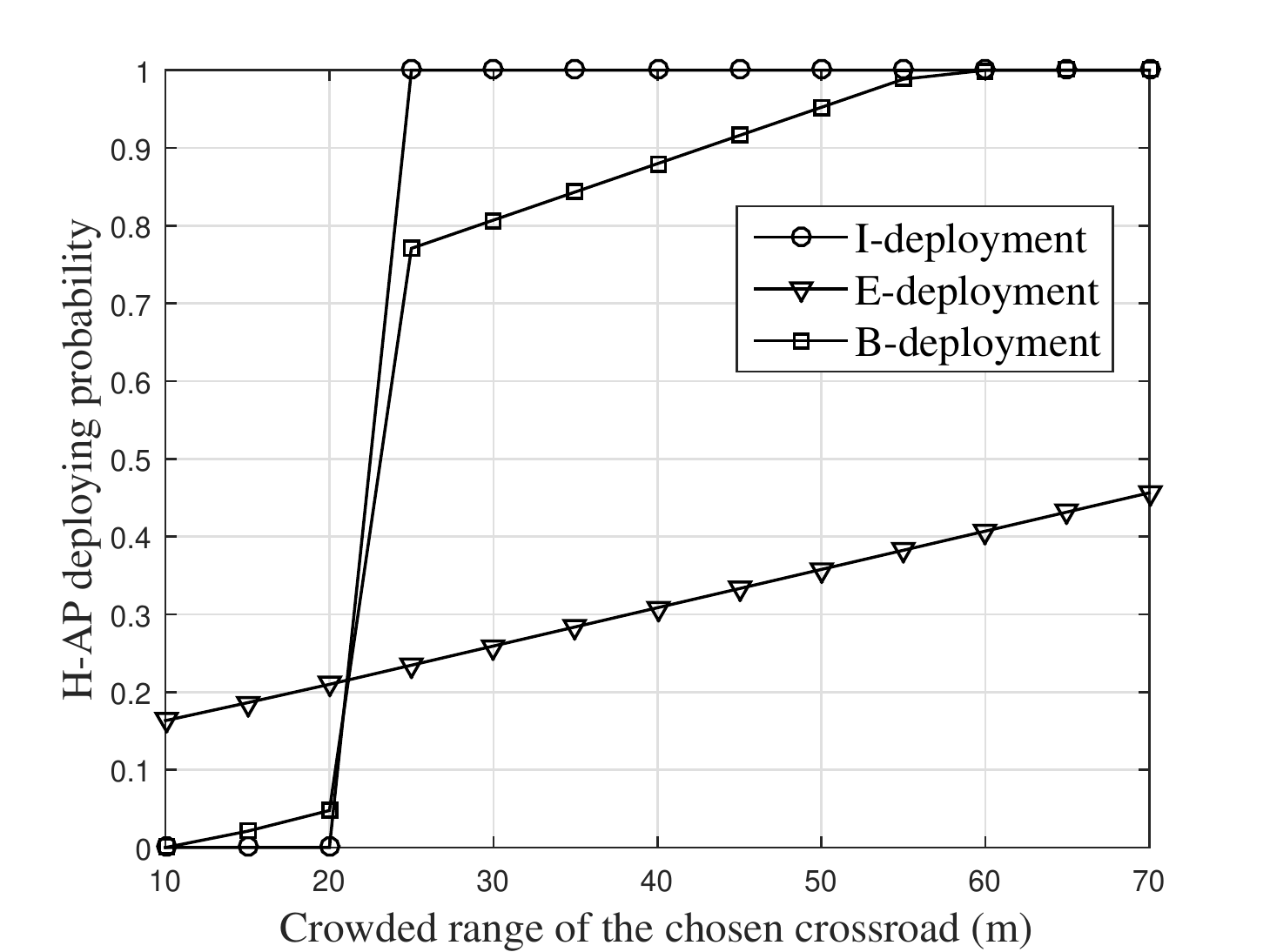}
  \caption{H-AP deploying probabilities versus crowded ranges}\label{probabilities}
\end{figure}

We randomly choose a crossroad and change its crowded range for evaluating its impact on the users' sojourn duration and the probability of an H-AP being deployed at this crossroad, while the crowded ranges of other crossroads are fixed. Observe from Fig.\ref{rC} that, users' sojourn duration within the WIT range and the WET range of the  H-AP deployed at the chosen crossroad both firstly increase and then reduce as its crowded range increases. Furthermore, when the crowded range is the same as the WIT range $r^{D}$ (or the same as the WET range $r^{E}$), the sojourn duration within the WIT range (or the WET range) is maximised.  By contrast, the user's sojourn duration within the WIT range and that within the WET range of the H-AP deployed at another crossroad both reduces as we increase the crowded range of the chosen crossroad. These observations are in line with Remark 1.

Furthermore,  we  plots the probabilities of an H-AP being deployed at the chosen crossroad against its crowded range in Fig.\ref{probabilities}. Observe from Fig.\ref{probabilities} that the probability of the H-AP being deployed is a non-decreasing function with respect to the crowded range of the chosen crossroad, regardless of any specific deployment scheme.  Moreover, the H-AP's deploying probability in the I-deployment scheme is either 0 or 1. Furthermore, the deploying probability of  the H-AP in the B-deployment scheme is a compromise between the I-deployment scheme and the E-deployment scheme, since it aims for balancing both the information downloading requirement and the energy harvesting requirement of the users.

\subsection{Information Downloading and Energy Harvesting}
\begin{figure}[!t]
  \centering
  \includegraphics[width=3in]{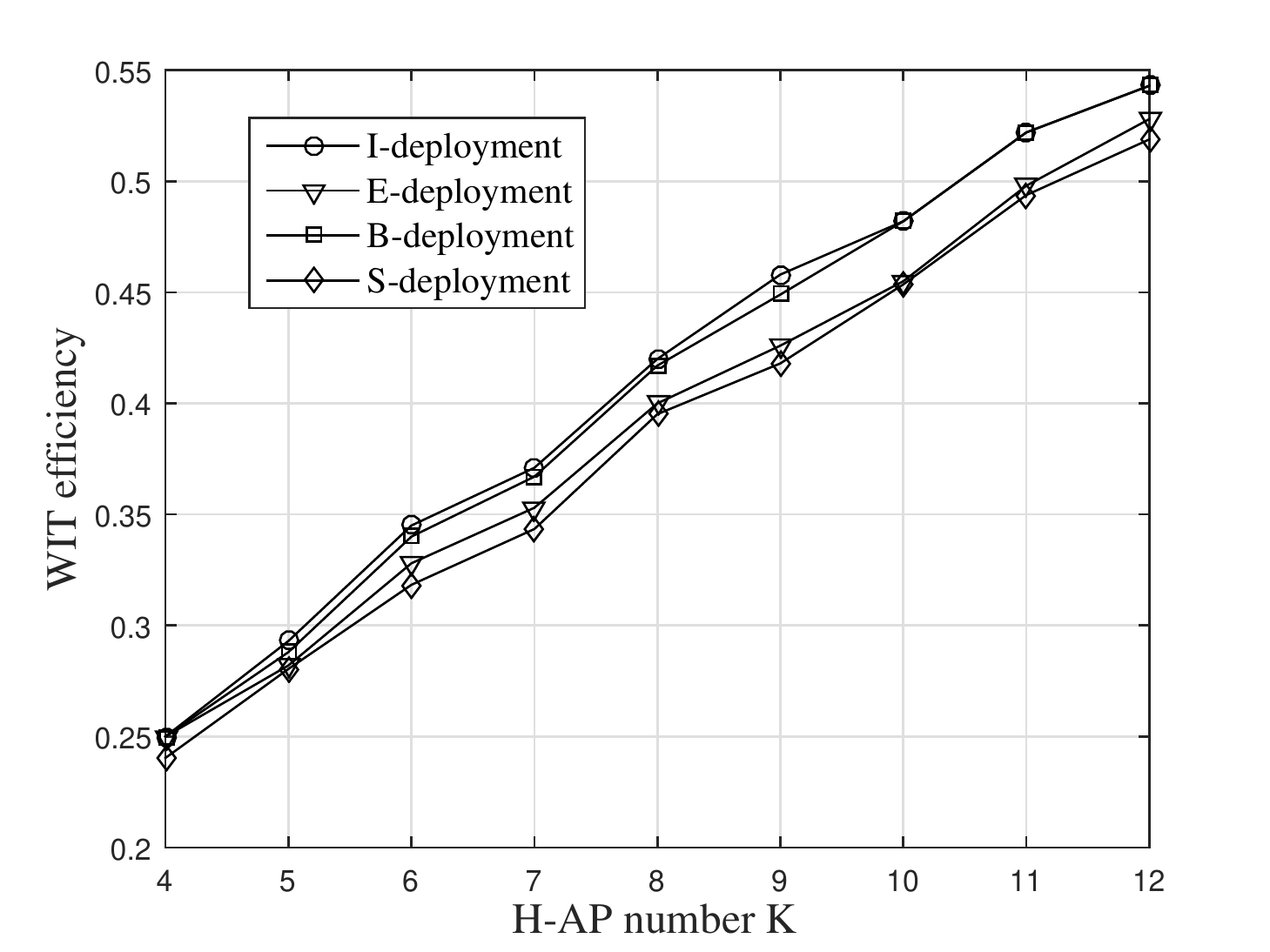}
  \caption{WIT efficiency versus number of H-APs}\label{dK}
\end{figure}
\begin{figure}[!t]
  \centering
  \includegraphics[width=3in]{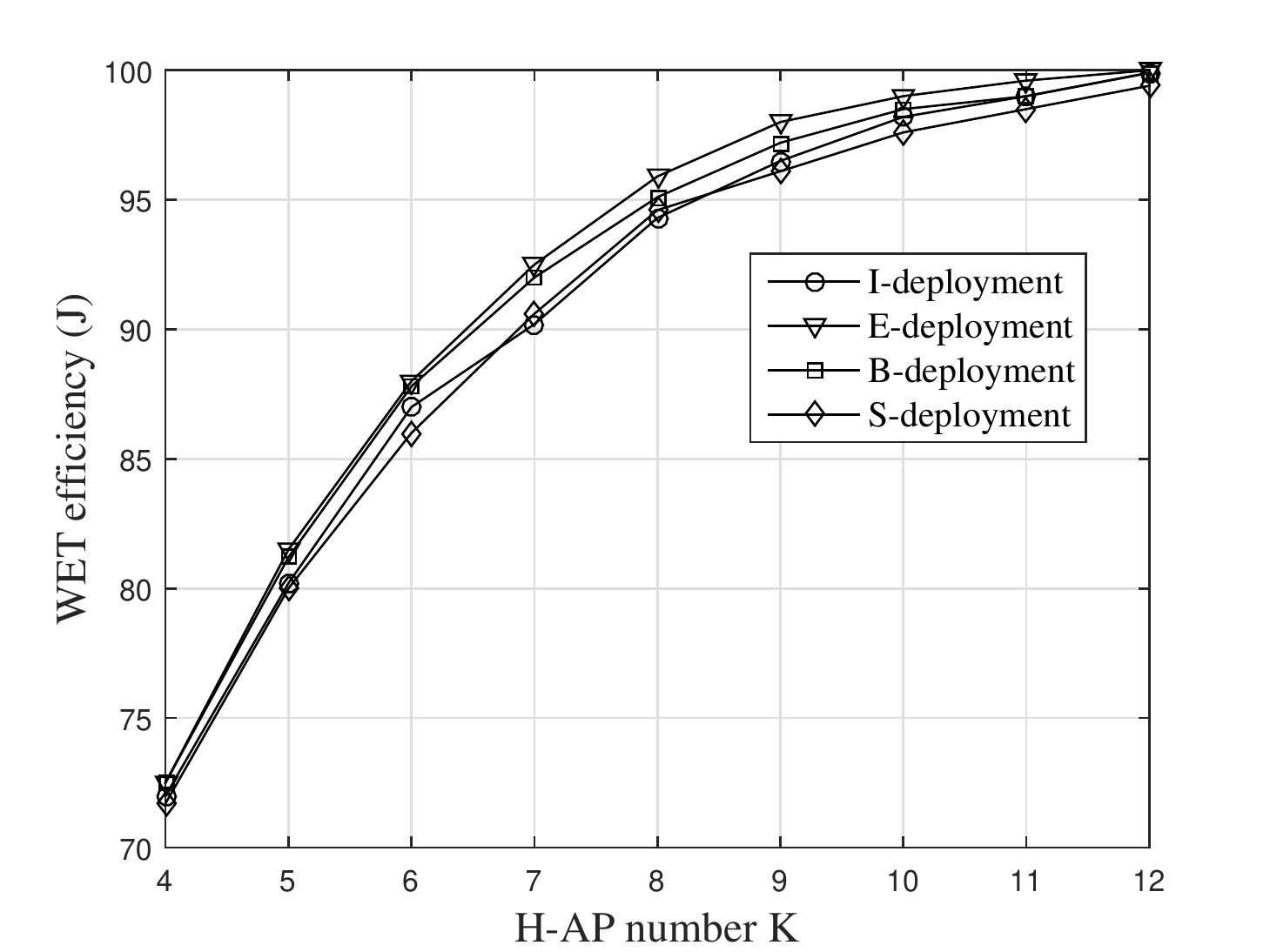}
  \caption{WET efficiency versus number of H-APs}\label{eK}
\end{figure}

 In Fig.\ref{dK} and Fig.\ref{eK}, we plot both the WIT and the WET efficiencies against the number of the H-APs pending to be deployed.  We choose the S-deployment scheme originally proposed by Ying \textit{et al} \cite{2626}. In this scheme, the H-APs are deployed by only considering the visiting frequency of users at each crossroad. We can readily obtain that our proposed H-AP deployment schemes are better than the S-deployment in terms of both the WET and WIT efficiencies.  Observe from Fig.\ref{dK} that the I-deployment scheme outperforms its counterparts in terms of the WIT efficiency, while the E-deployment scheme performs best in terms of the WET efficiency, as illustrated in Fig.\ref{eK}. Furthermore, as shown in both Fig.\ref{dK} and Fig.\ref{eK}, the B-deployment scheme achieves a compromise performance between the I-deployment scheme and the E-deployment one. Moreover, both the WIT and WET efficiencies increase,  as we increase the number of the H-APs.
\begin{figure}[!t]
  \centering
  \includegraphics[width=3in]{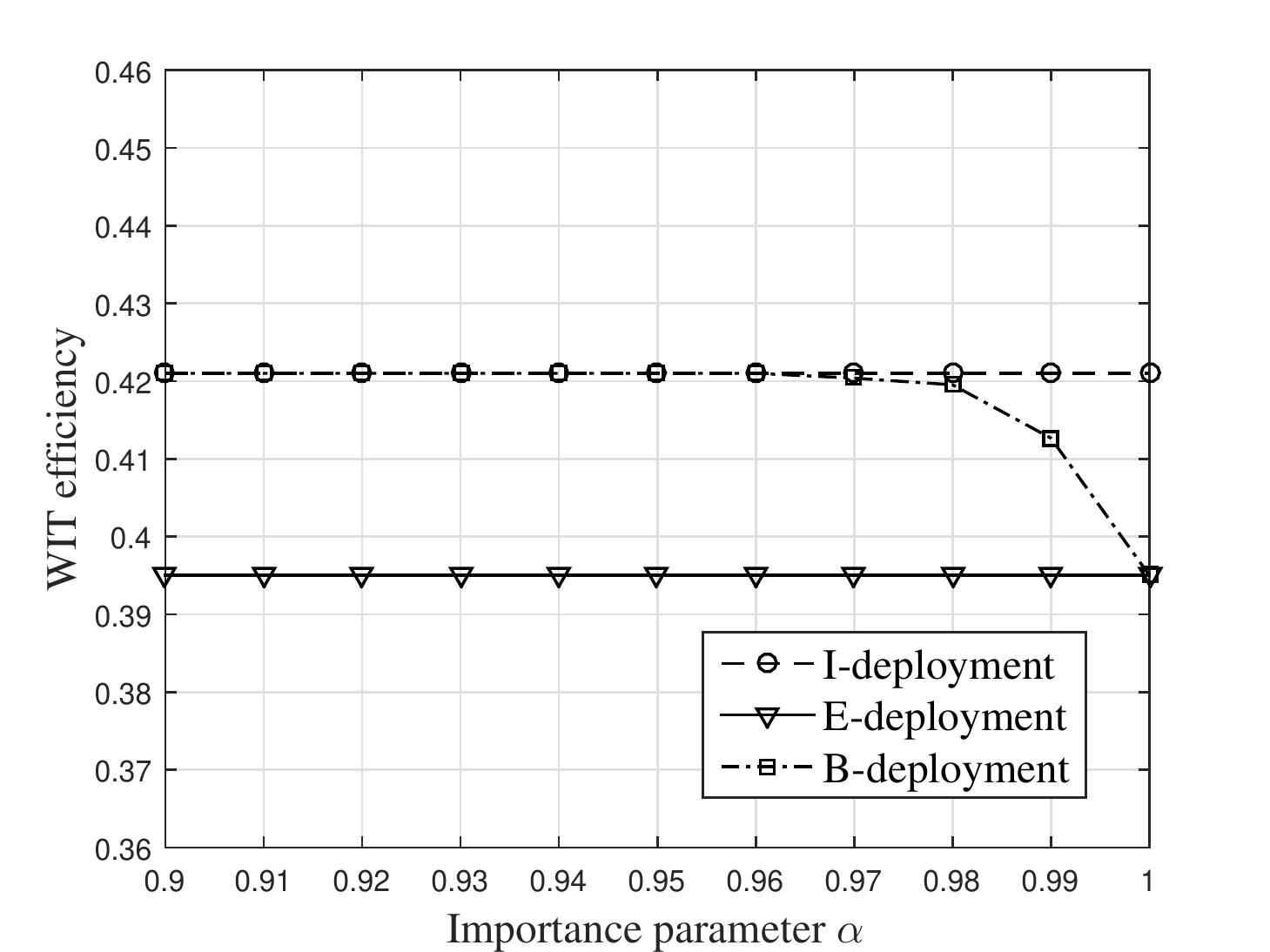}
  \caption{WIT efficiency versus importance parameter $\alpha$}\label{dalpha}
\end{figure}
\begin{figure}[!t]
  \centering
  \includegraphics[width=3in]{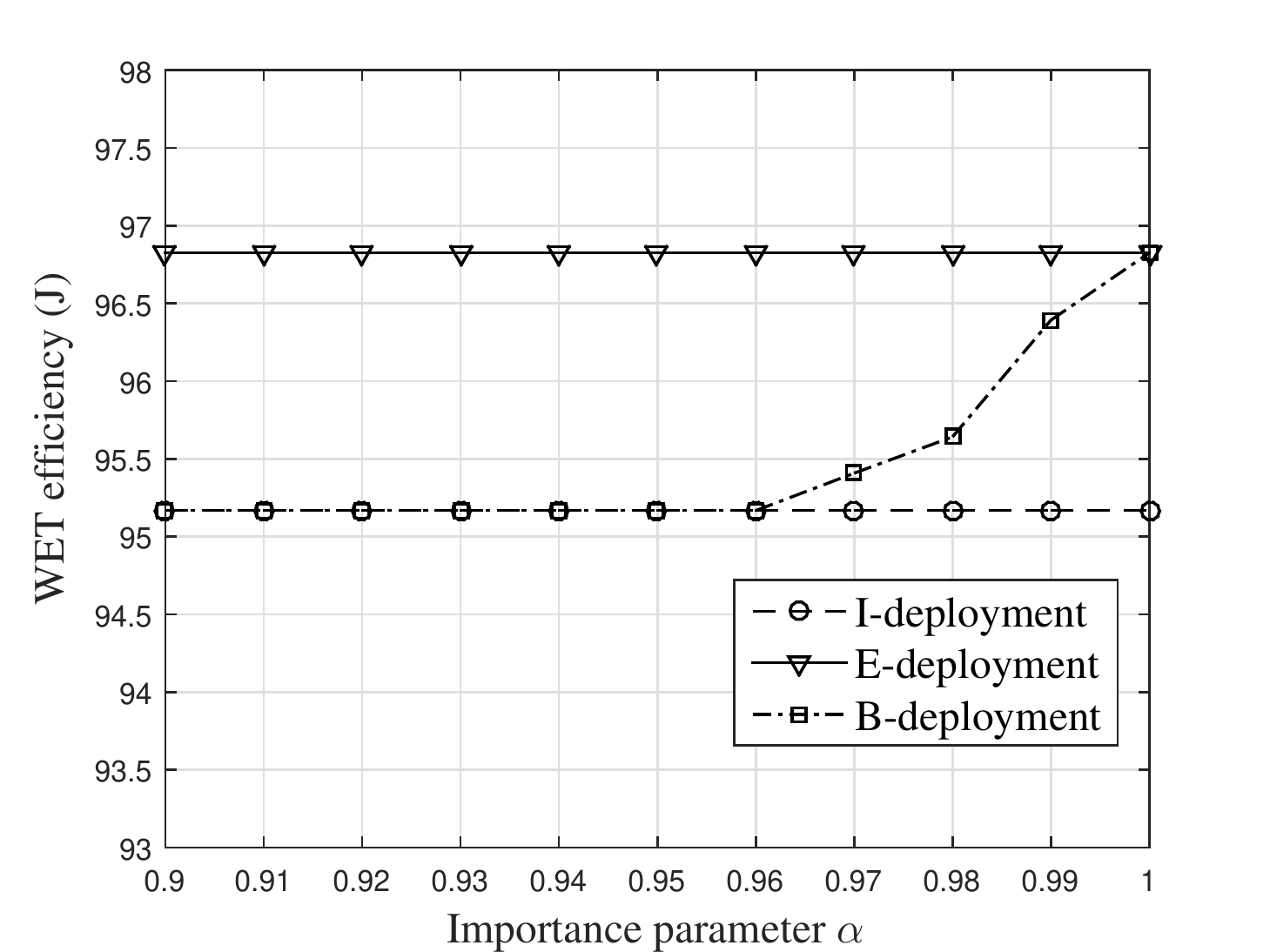}
  \caption{WET efficiency versus importance parameter $\alpha$}\label{ealpha}
\end{figure}

In Fig.\ref{dalpha} and Fig.\ref{ealpha}, we investigate the impact of the preference parameter $\alpha$ on the performance of the B-deployment scheme. Observe from Fig.\ref{dalpha} that the I-deployment performs best in terms of the information downloading performance. When $\alpha$ is lower than $0.95$, which indicates a relatively relaxed energy constraint, the B-deployment scheme has the same solution as the I-deployment scheme. By contrast, as $\alpha$ continuously grows, the information downloading performance of the B-deployment reduces until reaches that of the E-deployment. In Fig.\ref{ealpha}, the opposite trend is observed. As the preference parameter $\alpha$ continues to increase beyond $0.96$, the energy harvesting performance of the B-deployment scheme increases until it reaches that of the E-deployment.

\section{Conclusion}

The H-APs are deployed for the sake of simultaneously satisfying both the information downloading and the energy harvesting requirements of the users. These users move within the grid based city street blocks, whose mobility patterns are analysed by exploiting a two-dimensional Markov chain. According to the users' mobility patterns and the popularity (represented by the crowded range) of the crossroads, the B-deployment scheme of the H-APs has been proposed,  for striking a balance between the WIT and WET efficiencies. Both the theoretical and simulation results demonstrate that deploying the H-APs at the crowded crossroads may substantially improve both the WIT and the WET efficiencies. Additionally, changing the parameter $\alpha$ is capable of flexibly adjusting our preference on either the WIT or the WET efficiencies in the B-deployment scheme.

\appendices
\section{}
We will choose the analysis of information range for example, while that of energy range will be the same.

Let $s_{i_0,j_0}$ as our target crossroad, while the crowded range of other crossroads are fixed. When $r^C_{i_0,j_0}$ is increasing, there are two possible cases, namely $r^C_{i_0,j_0}<r^D$ and $r^C_{i_0,j_0}\ge r^D$. Hence, we will analyse the total sojourn duration of the each user in WIT range of a crossroad for these two cases, respectively.

\textbf{Cases 1:} $r^C_{i_0,j_0}<r^D$. With the aid of (2),(4) and (6), we may readily obtain the total sojourn duration of the user $u_m$ within the WIT range of the H-AP deployed at the crossroad $s_{i_0,j_0}$ as
\begin{align}
\tau^D_{m,i_0,j_0}&=\frac{\phi^m_{i_0,j_0}D^D_{m,i_0,j_0}}{\sum\limits_{i,j\neq i_0,j_0}\phi^m_{i,j}D^S_{m,i,j}+\phi^m_{i_0,j_0}D^S_{m,i_0,j_0}}\notag\\
&=1-\frac{\sum\limits_{i,j\neq i_0,j_0}\phi^m_{i,j}D^S_{m,i,j}+\phi^m_{i_0,j_0}(2r^D_{i_0,j_0}-l)/v_{m,0}}{\sum\limits_{i,j\neq i_0,j_0}\phi^m_{i,j}D^S_{m,i,j}+\phi^m_{i_0,j_0}D^S_{m,i_0,j_0}}\label{eq:appendixA1}
\end{align}

Observe from \eqref{eq:appendixA1} that, $r^C_{i_0,j_0}$ is only correlated with $D^S_{m,i_0,j_0}$ at the denominator of the second term in \eqref{eq:appendixA1}. Since $D^S_{m,i_0,j_0}$ is an monotonously increasing function with respect to  $r^C_{i_0,j_0}$ according to (2), $\tau^D_{m,i_0,j_0}$ is also a monotonously increasing function with respect to $r^C_{i_0,j_0}$.

By considering another crossroad  $s_{i_1,j_1}$, the total sojourn duration of the user $u_m$ within the WIT range  of the H-AP deployed at this crossroad can be expressed as
\begin{align}
\tau^D_{m,i_1,j_1}=\frac{\phi^m_{i_1,j_1}D^D_{m,i_1,j_1}}{\sum\limits_{i,j\neq i_0,j_0}\phi^m_{i,j}D^S_{m,i,j}+\phi^m_{i_0,j_0}D^S_{m,i_0,j_0}}
\end{align}
Since the crowded range $r^C_{i_1,j_1}$ of the crossroad $s_{i_1,j_1}$ does not change, $D^D_{m,i_1,j_1}$ is also unchanged. As a result, $\tau^D_{m,i_1,j_1}$ reduces as we increase $r^C_{i_0,j_0}$, since $D_{m,i_0,j_0}^S$ is a monotonously increasing function with respect to $r_{i_0,j_0}^{C}$.

\textbf{Cases 2:} $r^C_{i_0,j_0}\ge r^D$. The total sojourn duration of the user $u_m$ within the WIT range of  the H-AP deployed at the crossroad $s_{i_0,j_0}$ can be expressed as
\begin{align}
\tau^D_{m,i_0,j_0}=\frac{\phi^m_{i_0,j_0}D^D_{m,i_0,j_0}}{\sum\limits_{i,j\neq i_0,j_0}\phi^m_{i,j}D^S_{m,i,j}+\phi^m_{i_0,j_0}D^S_{m,i_0,j_0}}
\end{align}
Since we have $r^C_{i_0,j_0}\ge r^D$, $D^D_{m,i_0,j_0}$ is not correlated to $r^C_{i_0,j_0}$ anymore according to (4). Hence, $\tau^D_{m,i_0,j_0}$ reduces as we have an increasing $r^C_{i_0,j_0}$,  since $D^S_{m,i_0,j_0}$ is a monotonously increasing function with respect to $r_{i_0,j_0}^{C}$.

The sojourn duration $\tau^D_{m,i_1,j_1}$  of the user $u_m$ within the WIT range of the H-AP deployed at another crossroad $s_{i_1,j_1}$ is also a monotonously decreasing function with respect to $r_{i_0,j_0}^C$, according to the same analysis made in Case 1.

As a result, we have demonstrated that $\tau^D_{m,i_0,j_0}$ is a convex function with respect to $r_{i_0,j_0}^C$ and it can be maximised if we have $r_{i_0,j_0}^C = r^D$. Moreover, $\tau^D_{m,i_1,j_1}$ is a monotonously decreasing function with respect to $r_{i_0,j_0}^C$.

Furthermore, by exploiting the same methodology, we may also demonstrate the properties of the sojourn duration of a specific user within the WET range of the H-AP deployed at a specific crossroad.

\section{}
According to (6), the WIT efficiency of (7) can be re-formulated as
\begin{align}
\eta&=\frac{1}{M}\sum\limits_m\frac{\sum\limits_{i,j}\phi^m_{i,j}D^D_{m,i,j}y_{i,j}}{\sum\limits_{i,j}\phi^m_{i,j}D^S_{m,i,j}}\notag\\
&=\frac{1}{M}\sum\limits_m\eta_m
\end{align}

Then, we will analyse the monotonicity of $\eta$ in the following three cases.

\textbf{Case 1:} $y_{i_0,j_0}=1$ and $r^C_{i_0,j_0}<r^E$. In this case, $\eta_m$ can be re-formulated as
\begin{align}
\eta_m=\frac{\sum\limits_{i,j\neq i_0,j_0}\phi^m_{i,j}D^D_{m,i,j}y_{i,j}+\phi^m_{i_0,j_0}D^D_{m,i_0,j_0}}{\sum\limits_{i,j\neq i_0,j_0}\phi^m_{i,j}D^S_{m,i,j}+\phi^m_{i_0,j_0}D^S_{m,i_0,j_0}}\label{eq:etam}
\end{align}
Since we have $D^D_{m,i_0,j_0}=D^S_{m,i_0,j_0}-\frac{l-2r^D}{v_{m,0}}$ according to (2) and (4), we can  equivalently transform \eqref{eq:etam} as
\begin{align}
\eta_m=1-\frac{\sum\limits_{i,j\neq i_0,j_0}\phi^m_{i,j}(D^S_{m,i,j}-D^D_{m,i,j}y_{i,j})+\phi^m_{i_0,j_0}\frac{l-2r^D}{v_{m,0}}}{\sum\limits_{i,j\neq i_0,j_0}\phi^m_{i,j}D^S_{m,i,j}+\phi^m_{i_0,j_0}D^S_{m,i_0,j_0}}\label{eq:etam2}
\end{align}

Observe from \eqref{eq:etam2} that, since $D^S_{m,i_0,j_0}$ is a monotonously increasing function with respect to $r^C_{i_0,j_0}$, $\eta_m$ is also a  a monotonously increasing function with respect to $r_{i_0,j_0}^C$, and so is the total WIT efficiency $\eta$.

\textbf{Case 2:} $y_{i_0,j_0}=1$ and $r^C_{i_0,j_0}\ge r^E$. In this case, since $D^D_{m,i_0,j_0}$ is not correlated to $r^C_{i_0,j_0}$ anymore, and only $D^S_{m,i_0,j_0}$ in \eqref{eq:etam} is a monotonously increasing function with respect to $r^C_{i_0,j_0}$, $\eta_m$  is a monotonously decreasing function with respect to $r^C_{i_0,j_0}$ and so is the total WIT efficiency $\eta$.

\textbf{Case 3:} $y_{i_0,j_0}=0$. In this case, $\eta_m$ can be re-formulated as
\begin{align}
\eta_m=\frac{\sum\limits_{i,j\neq i_0,j_0}\phi^m_{i,j}D^D_{m,i,j}y_{i,j}}{\sum\limits_{i,j\neq i_0,j_0}\phi^m_{i,j}D^S_{m,i,j}+\phi^m_{i_0,j_0}D^S_{m,i_0,j_0}}
\end{align}
By exploiting the similar methodology, we can readily obtain that the total WIT efficiency $\eta$  is a monotonously decreasing function with respect to $r^C_{i_0,j_0}$.

To sum up, if no H-AP is deployed at the crossroad $s_{i_0,j_0}$, the total WIT efficiency $\eta$ reduces as the crowded range $r^C_{i_0,j_0}$ increases. By contrast, if a H-AP is deployed at the crossroad $s_{i_0,j_0}$, the total WIT efficiency $\eta$ is a convex function with respect to $r^C_{i_0,j_0}$, while it is maximised when we have  $r^C_{i_0,j_0}>r^D$.

\section{}
According to (2), (5) and (12), the total energy harvested by a single user $u_m$ can be re-formulated  as
\begin{align}
\varepsilon_m=\frac{\sum\limits_{i,j\neq i_0,j_0}\phi^m_{i,j}\varepsilon_{m,i,j}y_{i,j}+\phi^m_{i_0,j_0}\varepsilon_{m,i_0,j_0}y_{i_0,j_0}}{\sum\limits_{i,j\neq i_0,j_0}\phi^m_{i,j}D^S_{m,i,j}+\phi^m_{i_0,j_0}D^S_{m,i_0,j_0}}\label{eq:15}
\end{align}
When  we have $r^C_{i_0,j_0}>r^E$, the term $\varepsilon_{m,i_0,j_0}$ is not correlated to the crowded range $r^C_{i_0,j_0}$ anymore. According to (2),  $D^S_{m,i_0,j_0}$ in \eqref{eq:15} is a monotonously increasing function with respect to $r^C_{i_0,j_0}$. Therefore,  $\varepsilon_m$ is a monotonously decreasing function with respect to $r^C_{i_0,j_0}$. This remark has been proven.
\bibliography{VT-2017-02036}

\begin{IEEEbiography}[{\includegraphics[width=1in,height=1.25in,clip,keepaspectratio]{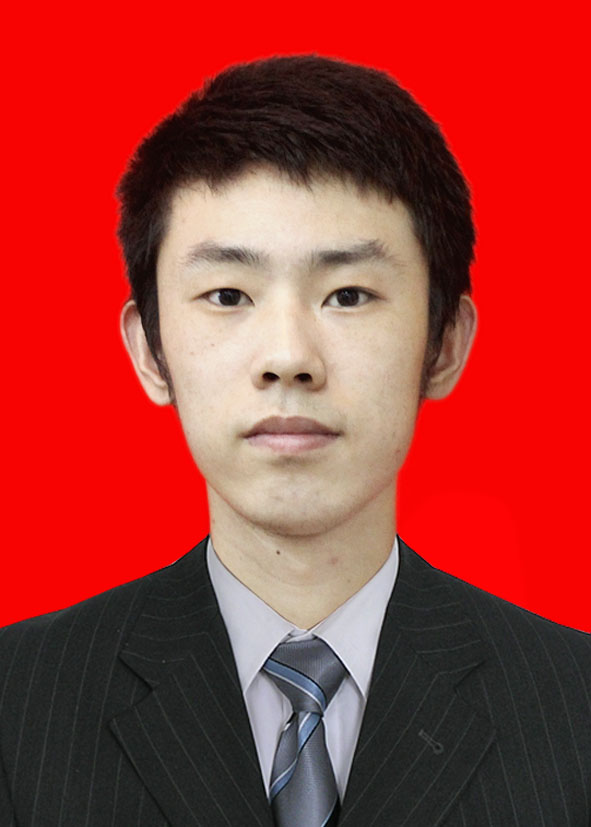}}]{Yizhe Zhao}
received the B.S. in communication engineering from Xidian University in 2014, and the M.S. in information and communication engineering from University of Electronic Science and Technology of China (UESTC) in 2017. He is currently pursuing the Ph.D. degree with the School of Information and Communication Engineering in UESTC. His research focuses on simultaneous wireless information and power transfer as well as data and energy integrated communication networks.
\end{IEEEbiography}

\begin{IEEEbiography}[{\includegraphics[width=1in,height=1.25in,clip,keepaspectratio]{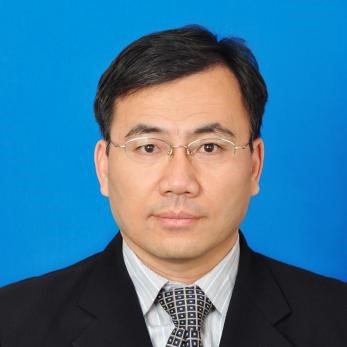}}]{Duohua Wang}
received the B.S. and M.S. degrees in communication engineering from Xidian University in 1998 and 2004, respectively. He is currently working at ZTE. His research focuses on wireless power transfer and radio access technologies.
\end{IEEEbiography}

\begin{IEEEbiography}[{\includegraphics[width=1in,height=1.25in,clip,keepaspectratio]{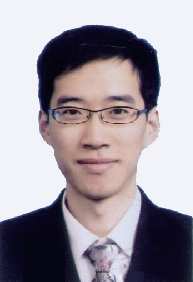}}]{Jie Hu}(S'11-M'16)
received his B.Eng. and M.Sc. degrees from Beijing University of Posts and Telecommunications, China, in 2008 and 2011, respectively, and received the Ph.D. degree from the Faculty of Physical Sciences and Engineering, University of Southampton, U.K., in 2015. Since March 2016, he has been working with the School of Information and Communication Engineering, University of Electronic Science and Technology of China (UESTC), China, as an Associate Professor. His research now is funded by National Natural Science Foundation of China (NSFC). He is also in great partnership with industry, such as Huawei and ZTE. He has served for ZTE Communications as the guest editor of the special issue "Wireless Data and Energy Integrated Communication Networks". He has a broad range of interests in wireless communication and networking, such as cognitive radio and cognitive networks, mobile social networks, data and energy integrated networks as well as communication and computation convergence.
\end{IEEEbiography}

\begin{IEEEbiography}[{\includegraphics[width=1in,height=1.25in,clip,keepaspectratio]{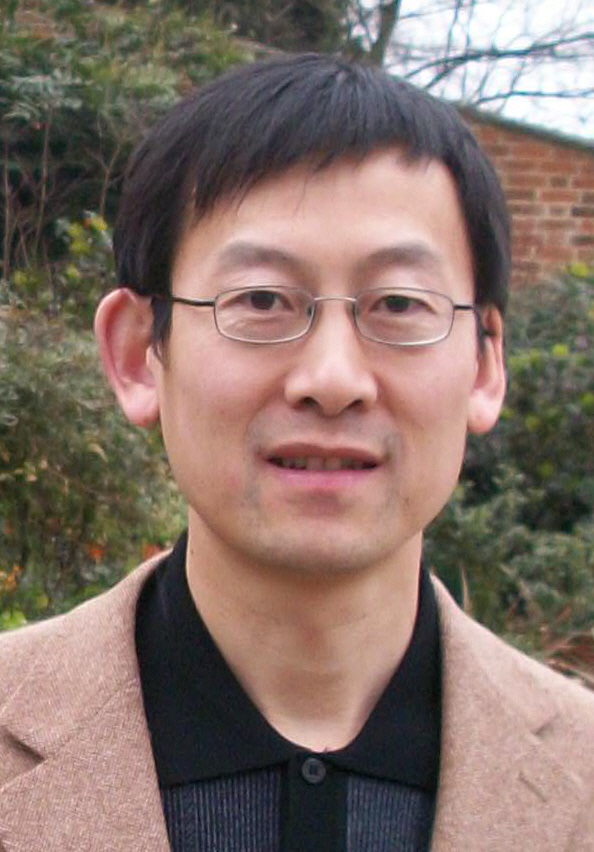}}]{Kun Yang}
received his PhD from the Department of Electronic and Electrical Engineering of University College London (UCL), United Kingdom. He is currently a full Professor in the School of Computer Science and Electronic Engineering, University of Essex, UK, heading the Network Convergence Laboratory. Before joining in University of Essex at 2003, he worked at UCL on several European Union (EU) research projects (such as FAIN, CONTEXT) for several years in the areas of IP network management, active networks and context-aware services. His main research interests include heterogeneous wireless networks, fixed mobile convergence, pervasive service engineering, future Internet technology and network virtualization, cloud computing. He manages research projects funded by various sources such as UK EPSRC, EU FP7 and industries. He has published 60+ journal papers in addition to 60+ conference papers. He is a Senior Member of IEEE and a Fellow of IET. He serves on the editorial boards of both IEEE and non-IEEE journals (Wiley and Springer) and have guest-edited several special issues in the above research areas. He has also served as (co-)chair (general or TPC) in many IEEE conferences.
\end{IEEEbiography}

\end{spacing}
\end{document}